\documentclass[a4paper,onecolumn,10pt]{IEEEtran}
\usepackage{graphicx}
\usepackage{subfigure,stfloats}
\usepackage{amssymb,amsthm}
\usepackage{tikz}

\usepackage[utf8]{inputenc} 
\usepackage[T1]{fontenc}
\usepackage{url}
\usepackage{ifthen}
\usepackage{cite}
\usepackage[cmex10]{amsmath} % Use the [cmex10] option to ensure complicance
\newtheorem{theorem}{Theorem}
\newtheorem{lemma}{Lemma}
\newtheorem{corollary}{Corollary}
\newtheorem{proposition}{Proposition}
\newtheorem{problem}{Problem}
\theoremstyle{remark}
\newtheorem{remark}{Remark}
\newtheorem*{example}{Example}

\newcommand{\wt}{\mathrm{wt}}

\newcommand\HH[1]{\frac12\mathrm{H}(#1,2)}

\def\0{\boldsymbol 0}
\def\1{\boldsymbol 1}
\def\2{\boldsymbol 2}
\def\3{\boldsymbol 3}

\def\PPP#1#2#3#4{P_{#2}^{#3}(#1,#4)}

\begin{document}
\title{On multifold packings of radius-$1$ balls\\ in Hamming graphs} 
\author{%
   \IEEEauthorblockN{Denis~S.~Krotov and Vladimir~N.~Potapov%
}\\
   \IEEEauthorblockA{Sobolev Institute of Mathematics,
                     Novosibirsk 630090, Russia\\
                     Email: \{krotov,vpotapov\}@math.nsc.ru}
}

\maketitle

\begin{abstract}\boldmath\bf
A $\lambda$-fold $r$-packing (multiple radius-$r$ covering) in a Hamming metric space is a code $C$ such that the radius-$r$ balls centered in $C$ cover each vertex of the space by not more (not less, respectively) than $\lambda$ times. The well-known $r$-error-correcting codes correspond to the case $\lambda=1$, while in general multifold $r$-packing are related with list decodable codes. We (a) propose asymptotic bounds for the maximum size of a $q$-ary $2$-fold $1$-packing as $q$ grows; (b) prove that a $q$-ary distance-$2$ MDS code of length $n$ is an optimal $n$-fold $1$-packing if $q\ge 2n$; (c) derive an upper bound for the size of a binary $\lambda$-fold $1$-packing and a lower bound for the size of a binary multiple radius-$1$ covering (the last bound allows to update the small-parameters table); (d) classify all optimal binary $2$-fold $1$-packings up to length $9$, in particular, establish the maximum size $96$ of a binary $2$-fold $1$-packing of length $9$; (e) prove some properties of $1$-perfect unitrades, which are a special case of $2$-fold $1$-packings. 
\end{abstract} 

%==!==!==!==!==!==!==!==!==!==!==!==!==!==!==!==!
%==!==!==!==!==!==!==!==!==!==!==!==!==!==!==!==!
%==!==!==!==!==!==!==!==!==!==!==!==!==!==!==!==!
\section{Introduction}%
It is well known that the error-correcting codes in Hamming spaces
can be treated as packings of balls of fixed radius $r$.
The balls in such packing do not intersect, 
every vertex of the space belongs to at most 
one of the balls, and this allows to correct an error
of weight at most $r$ if we know that the transmitted information 
was represented by a codeword. 
One of the main problems in the area is to find the maximum number
of balls in such packings, given the ball radius and the parameters 
of the metric space (for Hamming spaces, the alphabet size $q$ and the word length $n$).
We consider the generalization of this problem to the so-called 
$\lambda$-fold packings. 
For such packing, 
every vertex of the space belongs to at most 
$\lambda$ balls of the packing.
 In coding theory, the corresponding codes are known
as list decodable codes 
(for a good survey in the topic, see the introduction in~\cite{Elias:91}) 
and
can also be used in information transmission,
for example, in channels with noiseless feedback,
see~\cite{Ahlswede:73}.
% For example, in the case 
% of two-fold packing, if an error
% of weight at most $r$ occurred,
% then the receiver cannot uniquely determine the transmitted codeword;
% however, he has only two choices, and in the case of wrong choice,
% the transmitter can correct his choice by sending only one bit
% (such scheme implies that the channel has a noiseless feedback,
% so the transmitter can track what the receiver get;
% such channels can naturally occur in practice and 
% have been already considered in the literature, see e.g. \cite{Lebedev:2016}).
One of the main advantages of $\lambda$-fold 
%(or $\lambda$-fold, in general) 
packings is
that the cardinality of such a packing can be essentially larger than 
$\lambda$ times the cardinality of the largest one-fold packing in the same space.
In the current paper, we prove several bounds on the size
of $\lambda$-fold packings of radius-$1$ balls in binary and $q$-ary
Hamming schemes and consider properties of some special two-fold packings.%
\renewcommand\thefootnote{*}
\footnotetext{The work was funded by the Russian Science Foundation under grants 14-11-00555 
and 18-11-00136. The results of the paper were presented in part at the 
IEEE International Symposium on Information Theory ISIT 2019, July 7--12, 2019, Paris. }

The
 \emph{Hamming distance} $\mathrm{d_H}(x,y)$ 
 between two words
$x$ and $y$ of the same length 
is the number of coordinates in which
$x$ and $y$ differ.
The \emph{Hamming graph} (if $q=2$, the \emph{$n$-cube}) $H(n,q)$ 
is a graph whose vertices are the words of length $n$ over
the $q$-ary alphabet $\{0,\ldots,q-1\}$, 
two words being adjacent if and
only if they differ in exactly one position. 
The \emph{weight}
$\wt(x)$ of a word $x$ is the number of nonzeros in $x$.
The \emph{halved $n$-cube} $\HH{n}$ is a graph whose vertices are
the even-weight (or odd-weight) binary $n$-words, two words being
adjacent if and only if they differ in exactly two positions.
Two vertex sets of $H(n,q)$ are called \emph{equivalent} 
if some isomorphism of $H(n,q)$
sends one of the sets to the other.

We will say that a multiset $C$ of vertices of $H(n,q)$  is an
\emph{$\lambda$-fold $r$-packing} (of length $n$) 
if for every vertex $x$ of
$H(n,q)$ the number of elements of $C$ at distance at most $r$ from
$x$ does not exceed $\lambda$. 
The concept of $1$-fold $r$-packing
coincides with the well-known concept of $r$-error-correcting code.
The sphere-packing bound for error-correcting codes is generalized
to the obvious bound
\begin{equation}\label{eq:spb}
 \PPP n q \lambda r \le \lfloor \lambda q^n/|B_r| \rfloor
\end{equation}
on the maximum cardinality $\PPP n q \lambda r$ of a $\lambda$-fold $r$-packing in $H(n,q)$, where $B_r$ is
a radius-$r$ ball in $H(n,q)$.

A packing is called \emph{simple} 
if it has no multiple elements and can be treated as
an ordinary set, without multiplicities.
In the literature, the simple $\lambda$-fold $r$-packings are also known as the 
\emph{${\le}\lambda$-list decodable codes} with
radius $r$, see e.g. \cite{Blin86,AlonBP19}.
The only reason why we consider multisets is seeking generality:
some of our results do not require the simplicity of considered packings.

Preceding results on the bounds on the size of a $\lambda$-fold $r$-packing
(${\le}\lambda$\emph{-list decodable code})
were mainly focused on the asymptotics with growing packing radius. 
Blinovsky \cite{Blin86,Blin05} proved that  there exists a sharp
bound $\tau(\lambda,q)$ such that if $r=\tau n$,
$\tau<\tau(\lambda,q)$ then the maximum cardinality of a 
$\lambda$-fold $r$-packing  
is exponentially large in $n$;
in particular,
$\tau(\lambda,2)=\frac12-\frac{\binom{2k}{k}}{2^{2k+1}}$ where
$k=\lfloor \frac{\lambda}{2}\rfloor$.
In the case  $\tau>\tau(\lambda,q)$, the maximum
size of such packings  is bounded by constant as
$n\rightarrow\infty$ (see \cite{Blin09}). 
In~\cite{Blin86}, he also proved 
the asymptotic upper bound $\frac{\log|C|}{n}$
on the rate
 of a binary $\lambda$-fold 
$(\tau n)$-packings $C$. 
This bound was 
% generalized by Guruswami 
% and Vadhan~\cite{GurVad:2005} to an arbitrary 
% alphabet size and 
improved by 
Ashikhmin, Barg, and Litsyn~\cite{ABL00} 
for $\lambda=2$ 
and by Polyanskiy~\cite{Poly16} 
for $\lambda\geq 3$. 
How large can $C$ be when $\tau$ is just above the threshold 
$\tau(\lambda,q)$? 
For ordinary error-correcting codes ($\lambda = 1$),
the classical result of Levenshtein~\cite{Leven61} says
that the well-known Plotkin bound is sharp, 
namely, the size of the maximum code with radius
$(\tau(1,2)+\varepsilon)n=(\frac14+\varepsilon)n$ is
$\frac{1}{4\varepsilon}+O(1)$  as $\varepsilon\rightarrow 0$.
Alon et al.~\cite{AlonBP19} 
proved that the maximum possible size of a
$2$-fold packing with radius
$(\tau(2,2)+\varepsilon)n=(\frac14+\varepsilon)n$ is
$\Theta(\frac{1}{\varepsilon^{3/2}})$.
For $\lambda$-fold packings
with radius $(\tau(\lambda,2)+\varepsilon)n$, 
there is 
an upper bound
$O(\frac{1}{\varepsilon})$ 
for the size of $C$ as $\lambda$ is odd
\cite{AlonBP19}. 
In all mentioned asymptotic results, 
the length and the packing radius grow, 
while the alphabet is fixed. 

We are interested in other asymptotics and 
bounds for the size
of optimal multifold packings, where the packing radius $r$ 
is fixed and the length
or the size of the alphabet (or both) grows. In this paper, 
we focus on the case $r=1$, and obtain several different results,
separately considering the case of arbitrary alphabet
(Sections~\ref{s:q} and~\ref{s:mds}) with growing alphabet size
and the binary case (Sections~\ref{s:bounds}--\ref{s:uni}). Special attention
is payed to studying so-called $1$-perfect unitrades,
which are a special case of two-fold $1$-packings 
(where no radius-$1$ ball contains exactly one codeword).
The study of these objects is motivated by the connection with 
the class
of $1$-error-correcting perfect codes and by the fact that the classification 
of $1$-perfect unitrades of small parameters results in finding optimal
binary two-fold $1$-packing. 
Further studying $1$-perfect unitrades with theoretical and computer-aided 
tools can result if finding more good (may be even optimal) two-fold packings, 
not only in the binary case.

In Section~\ref{s:q} we propose asymptotic bounds 
for $2$-fold $1$-packing
in $H(n,q)$ as $q$ grows.
In Section~\ref{s:mds} 
we show that if $q \ge 2n$ 
then the maximum 
$n$-fold packings has size $q^{n-1}$;
an example of such packing is a distance-$2$ MDS code.
In Section~\ref{s:bounds}, based on linear programming,
we derive upper bounds on the size 
of a $\lambda$-fold $1$-packing in $H(n,2)$; as a corollary,
we also establish corresponding lower bounds on the size 
of a $\mu$-fold $1$-covering.
In Section~\ref{s:9}, 
we describe the optimal binary $2$-fold $1$-packings 
of length up to $9$ and their connection with completely regular codes
and discuss multifold packings attaining the sphere-packing bound.
In Section~\ref{s:uni}, we consider properties of $1$-perfect unitrades;
we estimate the minimum cardinality and the inner radius of a $1$-perfect unitrade,
show that all $1$-perfect unitrades up to the length $7$ are covered by systematic constructions,
and classify the $1$-perfect unitrades of length $9$ 
(to be exact, their length-$10$
extensions), which results in finding 
an optimal $2$-fold $1$-packing of length $9$ and size $96$.

% Some of the results of the paper 
% (Sections~\ref{s:bounds}--\ref{s:uni})
% concern only binary $\lambda$-fold $1$-packing. 
In the binary case, 
Sections~\ref{s:bounds}--\ref{s:uni} are focused on, 
it is often convenient to study 
the even-weight extensions
of the $\lambda$-fold $1$-packings 
instead of the original objects.
By this reason, 
in the end of the introduction we mention
a fundamental one-to-one correspondence, 
which generalizes the well-known correspondence 
between binary $r$-error-correcting codes 
and their extended versions.
A set of words is called \emph{even-weight} 
if each of its elements has even weight.
Given a multiset $C$ of binary words, its \emph{extension}
$\overline C$ is obtained from $C$ 
by appending the parity-check bit to all words:
$$\overline C = \{ 
  (x_1,\ldots,x_n,x_1+...+x_n) 
   \mid (x_1,\ldots,x_n)\in C \}.$$
The inverse operation, projection, or 
\emph{puncturing}, is removing the last symbol
from all words of $C$.
Another important operation, \emph{shortening},
consists of removing from $C$ the words whose last 
symbol is different from $0$ and then removing the 
last symbol $0$ from all remaining words.
   
\begin{proposition}\label{p:even}
A multiset $C$ of binary words is a $\lambda$-fold $r$-packing of length $n$ 
if and only if its extension $\overline C$ is an even-weight $\lambda$-fold $r$-packing of length $n+1$.
\end{proposition}

\begin{IEEEproof}
 It is easy to check that if every radius-$r$ ball in $H(n,2)$ contains
 at most $\lambda$ words from $C$, then every radius-$r$ ball in $H(n+1,2)$ contains
 at most $\lambda$ words from $\overline C$, and if some radius-$r$ ball in $H(n,2)$ contains more than $\lambda$ words from $C$, then some radius-$r$ ball in $H(n+1,2)$ contains more than $\lambda$ words from $\overline C$.
\end{IEEEproof}

%==!==!==!==!==!==!==!==!==!==!==!==!==!==!==!==!
%==!==!==!==!==!==!==!==!==!==!==!==!==!==!==!==!
%==!==!==!==!==!==!==!==!==!==!==!==!==!==!==!==!
\section{Two-fold $1$-packings in $q$-ary Hamming graph}\label{s:q}

In this section, we estimate the asymptotic of the maximum size of 
a two-fold $1$-packing in $H(n,q)$ as $q$ grows with constant $n$.
Then, we note that this also provides an estimation for 
the asymptotic of the maximum size of 
a $\lambda$-fold $1$-packing for every $\lambda$ between $2$ and $n$.

\begin{theorem}\label{th:q}
The maximum size $\PPP nq21$ of a $2$-fold $1$-packing in $H(n,q)$, $n\ge 3$,
satisfies 
\begin{equation}\label{eq:q}
q^{n-1-o(1)}\leq \PPP nq21 = o(q^{n-1}) , 
\end{equation}
as $q\rightarrow\infty$ and $n=\mathrm{const}$.
\end{theorem}
The case $n=2$ is not covered by this theorem, but considered in Corollary~\ref{cor1Pac},
see the next section: $\PPP 2q21=q$. The case $n=1$ is trivial: $\PPP 1q21=2$.
\begin{IEEEproof}
An \emph{$n$-uniform hypergraph} is a pair
$(V,E)$ 
from a set $V$, whose elements are called \emph{vertices},
and a set $E$ of $n$-subsets of $V$, called \emph{{}edges}.
Let  $f_n(m,v,e)$  be the maximum number of {}edges in an $n$-uniform
hypergraph on $m$ vertices such that the union of any $e$ {}edges 
has more than $v$ vertices. 

In \cite{AlonSha06}, it was shown that
% $$m^{k-o(1)}\leq f_n(m,3(n-k)+k+1,3)=o(m^k) \mbox{ for } 2\leq k<n.$$ 
\begin{equation}\label{eq:Alon}
 m^{\frac{3n-v+1}{2}-o(1)}\leq f_n(m,v,3)=o(m^{\frac{3n-v+1}{2}}) 
\end{equation}
if $2\leq \frac{3n-v+1}{2}<n$
(for $(n,v)=(3,6)$, 
it was proved earlier in \cite{Ruzsa76}).

\emph{Lower bound}. Let us consider an $n$-uniform hypergraph with $q$ vertices
and $f_n(q,n+\lambda+1,\lambda+1)$ {}edges 
such that the union of any $\lambda+1$ 
{}edges has more than $n+\lambda+1$ vertices.
From every {}edge, 
we construct a codeword of length $n$ 
by listing the elements 
of the {}edge in some order 
(the order itself is not important, 
one can take random, or lexicographic, or any else).
We state that the resulting code $C$ of size 
$f_n(q,n+\lambda+1,\lambda+1)$ is a $\lambda$-fold packing.
Indeed, if it is not, then there are $\lambda+1$ codewords 
$x_0$, \ldots, $x_\lambda$
at distance at most $1$ from some word $c$, 
the center of a ball. 
But $x_0$, \ldots, $x_\lambda$ contain at most $n+\lambda+1$ 
different symbols in total 
($n$ symbols of $c$, plus at most one 
unique symbol in each of $x_i$, $i=0,\ldots,\lambda$),
which contradicts the definition of $f_n$.
So, in the case $\lambda=2$ we see that
the lower bound in \eqref{eq:q}
follows from the lower bound in \eqref{eq:Alon}.

\emph{Upper bound}.
Given a $2$-fold $1$-packing $C$,
we first construct its subset $C'$, 
without multiple codewords, 
such that
$|C'|\ge |C|/2$ and there are no two adjacent codewords in $C'$.
We can always do so because, by the definition of 
a $2$-fold $1$-packing, every codeword of $C$
has at most one neighbor in $C$, 
while a codeword of multiplicity $2$ has no neighbors in $C$.
Next,
we define a hypergraph with $nq$ vertices
$(i,a)$,
$i\in \{1,\ldots,   n\}$,
$a\in \{0,\ldots, q-1\}$
and $|C'|$ {}edges
$\{(1,c_1),\ldots,(n,c_n):$ $(c_1,\ldots,c_n)\in C'\}$.

(*) We claim that \emph{the union of every three {}edges,
corresponding to some codewords $x$, $y$, $z$ of $C'$,
has at least $n+4$ vertices.} 
The proof of the claim is divided into three cases.

\begin{itemize}
\item 
 If the distance between some two of 
 $x$, $y$, $z$
is larger than $3$, then (*) is trivial.
\item 
If the distance between two of them, say $x$ and $y$,
is $3$, then
the union of the corresponding two {}edges has
$n+3$ vertices. If the {}edge corresponding to $z$
contains another vertex, then the union has 
$n+4$ vertices and (*) holds. Otherwise, in each position $z$ coincides with $x$ or $y$. Since $x$ and $y$ differ in only three positions, we see that $z$ is at distance $2$ from one of them and at distance $1$ from the other, which contradicts to the definition of $C'$.
\item 
There are two possibilities for three words to be
at distance $2$ from each other, 
without loss of generality.
One is $000...0$, $110...0$, $220...0$, 
and (*) holds in this case. 
The other is $1000...0$, $0100...0$, $0010...0$; 
such three words belong to the same radius-$1$ ball (centered in $0...0$), 
and hence cannot lie in $C'$ 
by the definition of a $2$-fold $1$-packing.
\end{itemize}

Finally, we see that the constructed hypergraph 
cannot have more than $f_n(nq,n+3,3)$ {}edges,
by the definition of $f_n$. Since it has $|C'|$ {}edges and $|C'| \ge |C|/2$, from \eqref{eq:Alon} 
we derive 
$|C|\le 2 f_n(nq,n+3,3) = o((nq)^{n-1}) = o(q^{n-1})$
as $q\to \infty$ and $n=\mathrm{const}$.
\end{IEEEproof}

Similar asymptotic estimation 
for the case $\lambda=n$ is rather
simple (in the next section, we find the exact value 
of $\PPP nqn1$  for $q\ge 2n$).

\begin{proposition}\label{p:n}
In the case $\lambda=n$, the maximum 
size $\PPP n q \lambda 1$ of a 
$\lambda$-fold $1$-packing in $H(n,q)$ 
admits the following bounds:
$$
q^{n-1}\leq \PPP nqn1 
\leq
\frac{q^{n}}{q-1+1/n}.
$$
\end{proposition}

\begin{IEEEproof} 
An example of $n$-fold $1$-packing  
is the distance-$2$ MDS code
$\{(x_1,\ldots,x_n)\,:\, x_1+\ldots+x_n \equiv 0 \bmod q\}$. 
Such code exists for any $n$ and $q$ and  its
cardinality  equals  $q^{n-1}$.

% A number of a pairs consisting of  codewords and  balls containing
%these codewords is $|C|(n(q-1)+1)$. But any ball with radius $1$
%contains not more than $n$ codewords.
Since $|B_1|=n(q-1)+1$, we conclude from (\ref{eq:spb}) that
$|C|(n(q-1)+1)\leq nq^n$.
\end{IEEEproof}

% It is easy to see that for $n=3$ the largest $1$-fold $1$-packing is
% the repetition code of size $q$; the largest $3$-fold
% $1$-packing is an MDS code of size $q^2$. 
For an arbitrary $n$, 
the largest $\lambda$-fold $1$-packing, $\lambda=2,\ldots,n$, 
has the cardinality between 
the cardinalities of largest $2$-fold and $n$-fold $1$-packings. Thus, we have the following.

\begin{theorem}\label{th:log}
If $\PPP nq\lambda1$ is the maximum cardinality 
of a $\lambda$-fold $1$-packing in $H(n,q)$, 
$\lambda \in \{ 2,\ldots,n \}$,
then
$$ \log \PPP nq\lambda1\simeq (n-1)\log q
\qquad
\mbox{as $q\to\infty$ and $n=\mathrm{const}$.}
$$
\end{theorem}
   
Another observation which can be made
from Theorem~\ref{th:q} and Proposition~\ref{p:n} is that $\PPP nq21=o(q^{n-1})$
while  $\PPP nqn1=\Omega(q^{n-1})$, as a function in $q$.
So, the answer to the following question meets
$2<\lambda_n\le n$.
   
\begin{problem}
 What is the value $\lambda_n$ such that 
$\PPP nq{\lambda_n-1}1=o(q^{n-1})$ but
$\PPP nq{\lambda_n}1=\Omega(q^{n-1})$?
\end{problem}

% [!!!!!customize the notations in accordance to our context]
Brown, Erd\H{o}s, and S\'os \cite{BES73} conjectured that
$f_n(q,l(n-k)+k+1,l) =o(q^{k})$ 
as $q\rightarrow\infty$ and 
$2\leq k<n$, $l>2$. In the special case $k=n-1$, their conjecture implies that
$f_n(q, n+l, l)=o(q^{n-1})$ as $q\rightarrow\infty$ and $l>2$.
The arguments in the paragraph ``Lower bound'' of the proof of Theorem~\ref{th:q} (where $l=\lambda+1$)
shows that the conjecture is true for every $l \le \lambda_n$;
so, finding $\lambda_n$
implies solving the conjecture for all smaller values of $l$.
However, the inverse connection is not so clear, as the ``Upper bound''
arguments  in the proof of Theorem~\ref{th:q} are not generalized to an
arbitrary $\lambda$.

%==!==!==!==!==!==!==!==!==!==!==!==!==!==!==!==!
%==!==!==!==!==!==!==!==!==!==!==!==!==!==!==!==!
%==!==!==!==!==!==!==!==!==!==!==!==!==!==!==!==!
\section{Distance-$2$ MDS codes are optimal}\label{s:mds}

A \emph{distance-$d$ MDS code} is a set of vertices of $H(n,q)$ with
cardinality $q^{n-d+1}$ and minimum distance $d$ between codewords.
Distance-$2$ MDS codes exist for any $n,q\ge 2$, for example, 
$\{(x_1,\ldots,x_n)\,:\, x_1+\ldots+x_n \equiv 0 \bmod q\}$.
As can be seen from Proposition~\ref{p:n},
the distance-$2$ MDS codes
are asymptotically optimal $n$-fold $1$-packings as $q$ grows. 
In this section, we prove a general upper bound on the size of a $\lambda$-fold $1$-packing
in any regular graph, which in particular shows that the distance-$2$ MDS codes
are indeed largest $n$-fold $1$-packings if $q\ge 2n$. 
We conjecture that this also holds for $q \ge n$.

\begin{theorem}\label{thPac}
Let 
$G=(V,E)$ be a $r$-regular graph (or multigraph, with multiple edges and/or loops),
and let $\alpha$ be a nonnegative constant such that 
$|\theta|\geq \alpha$
for any eigenvalue $\theta$ of $G$.
If $C$ is a multiset of vertices of $G$
such that every vertex from $V$ is adjacent to at most
$\lambda$ elements of $C$ 
(taking into account the multiplicity of the edges in the case of multigraph),
 then $\displaystyle\frac{|C|}{|V|}\leq \frac
 {r\lambda-\alpha^2}{r^2-\alpha^2}$.
\end{theorem}

Before proving the theorem, we derive one simple inequality.
\begin{lemma}\label{pr1Pac}
Assume that a vector $\bar v =(v_1,\ldots,v_N)$ from $\mathbb{R}^N$
is orthogonal to the all-one vector ${\overline 1}$
(that is, $v_1 + \ldots + v_N=0$).
If $a$ and $b$ are nonnegative constants such that $-a\leq v_i\leq b$ 
for every $i$ from $\{1,\ldots,N\}$, then
$$\|\bar v\|^2 \leq a b N,$$
where $\|\bar v\|^2 = v_1^2+\ldots+v_N^2$; moreover, $\|\bar v\|^2 = a b N$
if and only if $v_i\in \{-a,b\}$, $i=1,\ldots,N$.
\end{lemma}
\begin{IEEEproof}
Consider the vector $\bar u=(u_1,\ldots,u_N)=\bar v+\frac{a-b}{2}{\overline 1}$. 
Straightforwardly,
$|u_i|\leq \frac{a+b}{2}$ for $i=1,\ldots,N$. 
Consequently,
$\|\bar u\|^2\leq\frac{(a+b)^2N}{4}$, 
with equality only if all components are $\pm\frac{a+b}{2}$.
From the orthogonality of $\bar v$ and 
${\overline 1}$,
we obtain $\|\bar v\|^2=\|\bar u\|^2-\|\frac{a-b}{2}{\overline 1}\|^2\leq\frac{(a+b)^2N}{4}-\frac{(a-b)^2N}{4}=a b N$,
with equality only if all components of $\bar v$ are in $\{-a,b\}$.
\end{IEEEproof}

\begin{IEEEproof}[Proof of Theorem~\ref{thPac}]
We consider the space of functions $f:V\to \mathbb{R}$, treated as 
$|V|$-tuples whose coordinates are indexed by the vertices from $V$.
The graph $G$ can be represented by the \emph{adjacency} 
matrix $M$ of size
$|V|\times |V|$ whose element $M_{a,b}$, $a,b\in V$,
equals the multiplicity of the edge $\{a,b\}$ in the multigraph.
The eigenvalues
$\theta_0, \ldots, \theta_D$ of the adjacency matrix are also known
as the eigenvalues of the graph.
By $\chi_C:V\to \{0,1,2, \ldots\}$, 
we denote the multiplicity vector 
of a vertex multiset $C$. In particular, $\overline 1 = \chi_V$.
Denote $\rho=\frac{|C|}{|V|}$ and $F=\chi_C - \rho{\overline 1}$.
So, $\chi_C=\rho{\overline 1}+F$,
where $F \perp {\overline 1}$. 
Consequently,
\begin{equation}\label{eq:FV}
\|F\|^2=\|\chi_C\|^2-\|\rho{\overline 1}\|^2=\rho(1-\rho)|V|.
\end{equation}
Consider the vector 
$\bar v=(v_x)_{x \in V} =M\chi_C=r\rho{\overline 1}+MF$. 
Since, by the hypothesis of the theorem, 
every vertex $x$ from $V$ has at most $\lambda$ neighbors in $C$,
we have $0\leq v_x\leq \lambda$,
and hence 
\begin{IEEEeqnarray*}{c}
-r\rho\leq (MF)_x\leq \lambda-r\rho
\end{IEEEeqnarray*}
(here the right expression is positive because
$\rho \leq \lambda/(r+1)$ by the sphere-packing bound).
From the regularity of the graph and $F \perp {\overline 1}$,
we find $(MF,\overline 1)=(F,M\overline 1)=r(F,\overline 1)=0$,
where $(\cdot,\cdot)$ is the inner product.
So, $MF \perp {\overline 1}$, and by  Lemma~\ref{pr1Pac} we get 
\begin{equation}\label{eq:ab}
\|MF\|^2 \le r\rho(\lambda-r\rho)|V|.
\end{equation}
Consider the representation
$F=f_0+\ldots+f_D$ of $F$ as the sum of 
eigenvectors $f_i$ of $M$ 
corresponding to the eigenvalues $\theta_i$, $i=0,\ldots,D$.
 Since $MF=Mf_0+\ldots+Mf_D=\theta_0f_0+\ldots+\theta_Df_D$ 
 and the vectors
 $f_i$ are pairwise orthogonal, 
 we  find 
\begin{equation}\label{eq:MFF}
 \|MF\|^2=\theta^2_0\|f_0\|^2+\ldots+
 \theta^2_D\|f_D\|^2\geq \alpha^2(\|f_0\|^2+\ldots+
 \|f_D\|^2)=\alpha^2 \|F\|^2.
\end{equation}
Finally, we conclude that
$$
\alpha^2\rho(1-\rho)|V|
\;\stackrel{\eqref{eq:FV}}=\;
\alpha^2\|F\|^2
\;\stackrel{\eqref{eq:MFF}}\leq\;
\|MF\|^2
\;\stackrel{\eqref{eq:ab}}\leq\;
r\rho(\lambda-r\rho)|V|.$$
So, $\alpha^2(1-\rho) \le r(\lambda-r\rho)$,
and
$\frac{|C|}{|V(G)|}=\rho\leq \frac
 {r\lambda-\alpha^2}{r^2-\alpha^2}$.
\end{IEEEproof}

\begin{corollary}\label{cor1Pac}
If $q\geq 2n$, 
then the maximum cardinality $\PPP nqn1$ 
 of an $n$-fold $1$-packing in $H(n,q)$
 equals $q^{n-1}$ and every $n$-fold $1$-packing in $H(n,q)$
 of cardinality $q^{n-1}$ is a distance-$2$ (MDS) code.
\end{corollary}
\begin{IEEEproof}
The eigenvalues of $H(n,q)$ are $\theta_i=-n+qi$, $i=0,\ldots,n$ 
\cite{Brouwer}.
If $q\geq 2n$, then obviously $|\theta_i|\ge n$ for every $i$. 
The graph $H(n,q)$ is $n(q-1)$-regular, so by Theorem~\ref{thPac}
every $n$-fold $1$-packing $C$ satisfies 
$$
\frac{|C|}{q^n}\leq \frac
 {n^2(q-1)-n^2}{n^2(q-1)^2-n^2}=\frac
 {(q-1)-1}{(q-1)^2-1}=\frac{1}{q}.
 $$
On the other hand, any distance-$2$ MDS code $C$ in $H(n,q)$,
for example, 
$$
C=\{(x_1,\ldots,x_n)\in\{0,\ldots,q-1\}^n \mid x_1+\ldots+x_n =0\bmod q \},
$$
is an $n$-fold
 $1$-packing that meets 
 $\frac{|C|}{q^n}=
\frac{1}{q}$.

It remains to show that every such packing $C$ is a distance-$2$ code. 
As follows from Lemma~\ref{pr1Pac},  \eqref{eq:ab} holds with equality 
only if every vertex has exactly $0$ or $\lambda$ neighbors from $C$.
But, by the definition of a $\lambda$-fold $1$-packing,
any codeword of $C$ has at most $\lambda-1$ neighbors from $C$; 
hence, it has $0$ neighbors from $C$. 
So, the minimum distance between different codewords is at least $2$.
\end{IEEEproof}

If $q=n-1$ and $q$ is a prime power, 
then for any $\lambda$ from $\{1,\ldots,q^2\}$ we can construct 
a $\lambda$-fold $1$-packing in $H(n,q)$ 
as the union of $\lambda$ cosets of the $1$-error-correcting
Hamming code of cardinality $q^{n-2}$.
Such packing has cardinality 
$\lambda q^{n-2}=\frac{\lambda q^n}{nq-q+1}$, 
which is larger than
$\frac{\lambda q^n}{nq}$;
in particular, if $\lambda=n$, then it is larger than the cardinality
$q^{n-1}$ of a distance-$2$ MDS code.

  {\bf
Conjecture.} If $q\geq n$, then the cardinality of the largest $n$-fold $1$-packing in $H(n,q)$
 is $q^{n-1}$.

%==!==!==!==!==!==!==!==!==!==!==!==!==!==!==!==!
%==!==!==!==!==!==!==!==!==!==!==!==!==!==!==!==!
%==!==!==!==!==!==!==!==!==!==!==!==!==!==!==!==!
\section{Bounds for multifold packings 
and multiple coverings in the binary case}\label{s:bounds}

In this section, we prove upper bounds on the size of 
$\lambda$-fold $1$-packings in the $n$-cube $H(n,2)$,
generalizing the bounds and approach from \cite{BesBro77}
for $1$-error-correcting codes, corresponding to the case 
$\lambda=1$. As a corollary, based on the connection between 
multifold packings and multiple coverings 
(see the definition in Subsection~\ref{s:covering}),
new lower bounds for binary multiple radius-$1$ coverings
are established.

\subsection{Multiple $1$-packings}\label{s:packing}
 The \emph{weight distribution} of a code $C$ of length $n$ is the
sequence $\{A_i\}_{i=0}^n$, where $A_i$ is the number of the
codewords of weight $i$ in $C$. 
The weight distribution $\{A_i(x)\}_{i=0}^n$ of $C$ with
respect to a word $x$ is the weight distribution of the code $C+x$.
The \emph{distance distribution} $\{B_i\}_{i=0}^n$ of $C$ is defined as the average weight
distribution of $C$ with respect to all its codewords: $B_i=\frac{1}{|C|}\sum_{x \in C}A_i(x)$. The main result of this section is the following bound.

\begin{theorem}
\label{th:main1_} The maximum size $\PPP n2\lambda 1$ of a 
binary $\lambda$-fold $1$-packing of
length $n$, where  $\lambda\equiv \sigma \bmod 2$,
$\sigma\in\{0,1\}$, satisfy
\begin{enumerate}
 \item[\rm (a)] 
 $\displaystyle \PPP n2\lambda 1 \leq \frac{2^{n}(\lambda n+3\lambda-4+\sigma)}{n(n+4)}$
 \qquad \mbox{if $n\equiv 0 \bmod 4$}, \\
 \item[\rm (b)] 
 $\displaystyle \PPP n2\lambda 1 \leq \frac{2^{n}(\lambda n+\lambda-2)}{(n-1)(n+3)}$
 \qquad \mbox{if $n\equiv 1 \bmod 4$}, \\
 \item[\rm (c)] 
 $\displaystyle \PPP n2\lambda 1 \leq \frac{2^{n}(\lambda n+\lambda-2+\sigma)}{n(n+2)}$
 \qquad \mbox{if $n\equiv 2 \bmod 4$}, \\
 \item[\rm (d)] 
 $\displaystyle \PPP n2\lambda 1 \leq \frac{2^{n}\lambda}{n+1}$
 \qquad \mbox{if $n\equiv 3 \bmod 4$}.
 \end{enumerate}
\end{theorem}

We will prove this theorem in the following form, whose claim is equivalent 
to the claim of Theorem~\ref{th:main1_} by Proposition~\ref{p:even}.

\begin{theorem}
\label{th:main1} Every even-weight $\lambda$-fold $1$-packing $C$ of
length $n$, where  $\lambda\equiv \sigma \bmod 2$,
$\sigma\in\{0,1\}$, satisfy
\begin{enumerate}
 \item[\rm (a)] $\displaystyle |C| \leq \frac{2^{n-1}(\lambda n+2\lambda-4+\sigma)}{(n-1)(n+3)}$\qquad \mbox{if $n\equiv 1 \bmod 4$}, \\
 \item[\rm (b)] $\displaystyle |C| \leq \frac{2^{n-1}(\lambda n-2)}{(n-2)(n+2)}$\qquad \mbox{if $n\equiv 2 \bmod 4$}, \\
 \item[\rm (c)] $\displaystyle |C| \leq \frac{2^{n-1}(\lambda n-2+\sigma)}{(n-1)(n+1)}$\qquad \mbox{if $n\equiv 3 \bmod 4$}, \\
 \item[\rm (d)] $\displaystyle |C| \leq \frac{2^{n-1}\lambda}{n}$\qquad \mbox{if $n\equiv 0 \bmod 4$}.
 \end{enumerate}
\end{theorem}

\begin{IEEEproof}
Let $\{B'_i\}_{i=0}^n$ be the
MacWilliams transform of the distance distribution $\{B_i\}_{i=0}^n$ of $C$; that is,
\begin{IEEEeqnarray}{rCl}
\label{eq:B=A}
|C| B'_k & = & \sum_{i=0}^n B_i K_k(i),\nonumber\\
\label{eq:B=A'} 2^nB_k & = & |C|\sum_{i=0}^n B'_i K_k(i),
\qquad k=0,\ldots,n,
\end{IEEEeqnarray}
where
\[
K_k(i) = \sum_{j=0}^k(-1)^j {\binom{i}{j}}{\binom{n-i}{k-j}}
\]
is a Krawtchouk polynomial; in particular,
\begin{IEEEeqnarray*}{rClrCl}
K_0(i) & = & 1, &
K_2(i) & = & \frac12(n-2i)^2 - \frac12 n, \\
K_{n-1}(i) &=& (-1)^i(n-2i), \quad&
K_{n}(i) &=&(-1)^i.
\end{IEEEeqnarray*}
It is well known that $B'_0 = 1$ and $B'_i \geq 0$ for $1 \leq i
\leq n$ \cite{Delsarte:1973}.

As $C$ is an even-distance code, $B_i = 0$ for odd $i$, and, since
$K_{n-k}(i) = (-1)^{i} K_{k}(i)$, we have
\begin{equation}
\label{eq:mirror} B'_{k} = B'_{n-k}.
\end{equation}

(a) Let $n\equiv 1 \bmod 4$. Define $\alpha(i) = (n-3)
K_0(i)+2K_2(i)+2K_{n-1}(i)$. Direct calculations now show that
\begin{equation}
\label{eq:defal} \alpha(i) = 
\left(n-2i-2+(-1)^i\right)\left(n-2i+2+(-1)^i\right).
\end{equation}
It turns to $(n-3-2i)(n+1-2i)$ with zeros in 
$\frac{n-3}2$, $\frac{n+1}2$ if $i$ is odd,
and to $(n-1-2i)(n+3-2i)$  with zeros in 
$\frac{n-1}2$, $\frac{n+3}2$ if $i$ is even.
From (\ref{eq:defal}) and $n \equiv 1 \pmod 4$ we derive
\begin{IEEEeqnarray}{rCll} \label{eq:4cases}
\alpha(i)&=&0 &\mbox{ for  
$i=\frac{n-3}{2}$, $\frac{n-1}{2}$,
$\frac{n+1}{2}$, $\frac{n+3}{2}$, }
\\ \nonumber
\alpha(i)&>&0 &\mbox{ for any other integer }i.
\end{IEEEeqnarray}
From the packing condition we have $B_{n-1} \leq \lambda$ and,
moreover,
\begin{equation}\label{eq:n-3a0+2a2}(n-3)B_0+2B_2 \le \lambda n-4+\sigma,\end{equation}
with equality only if there are no codewords of multiplicity more
than $1$.
\big(Indeed, for a vertex $x$ of multiplicity $A_0(x)=1$, 
the number $A_2(x)$ of codewords 
at distance $2$ from $x$ 
does not exceed
$\lfloor n(\lambda-1)/2\rfloor = (n(\lambda-1)+\sigma-1)/2$; 
so,
$$(n-3)A_0(x)+2A_2(x) \le n-3 + n(\lambda-1)+\sigma-1 = \lambda n-4+\sigma.$$ 
For a larger
multiplicity, $A_0(x)\ge 2$, we have 
\begin{IEEEeqnarray*}{rCl}
 (n-3)A_0(x)+2A_2(x) &\le& (n-3)A_0(x) +
n(\lambda - A_0(x)) \\ &\le& \lambda n - 6,
\end{IEEEeqnarray*}
which is smaller than the right part of~\eqref{eq:n-3a0+2a2}.\big)

Utilizing (\ref{eq:mirror}), we then get

\begin{IEEEeqnarray}{rCl}
\label{eq:al} 2\alpha(0)B'_0 & = & \alpha(0)B'_0 + \alpha(n)B'_n
\leq \sum_i \alpha(i) B'_i\nonumber \\
& = & \frac{2^n((n-3)B_0+2B_2+2B_{n-1})}{|C|}\\
&  \leq & \frac{2^n(\lambda n-4+\sigma+2\lambda )}{|C|} 
\nonumber
\end{IEEEeqnarray}

\noindent and thereby
\[
|C| \leq \frac{2^n(\lambda n+2\lambda-4+\sigma)}{2\alpha(0)B'_0} =
\frac{2^{n-1}(\lambda n+2\lambda-4+\sigma)}{(n-1)(n+3)}.
\]

(b) Let $n\equiv 2 \bmod 4$. Define $\beta(i) = (n-2)
K_0(i)+2K_2(i)-2K_{n}(i)$. Straightforwardly,
\begin{equation}\label{eq:defal2}
\beta(i) = (n-2i)^2-2-2(-1)^i.
\end{equation}
Since $n/2$ is odd,
we see that
$$
\beta(i) = 0,\qquad \mbox{if } i\in \{n/2-1,n/2,n/2+1\};
$$
for any other integer $i$, we have $\beta(i)>0$. From the packing
condition we have
$$(n-2)B_0+2B_2 \le \lambda n-2$$
with equality if and only if $B_0=1$ (i.e., there are no codewords
of multiplicity more than $1$) and $B_2=(\lambda-1)n/2$.
Then, we get

\begin{IEEEeqnarray}{rCl}
\label{eq:al2} 2\beta(0)B'_0 & = & \beta(0)B'_0 + \beta(n)B'_n
\leq \sum_i \beta(i) B'_i\nonumber \\
& = & \frac{2^n((n-2)B_0+2B_2-2B_{n})}{|C|}\\
&  \leq & \frac{2^n(\lambda n-2)}{|C|}\nonumber
\end{IEEEeqnarray}
and thereby
\[
|C| \leq \frac{2^n(\lambda n-2)}{2 \beta(0)B'_0} =
\frac{2^{n-1}(\lambda n-2)}{(n-2)(n+2)}.
\]

(c) Let $n\equiv 3 \bmod 4$. Define $\gamma(i) = (n-1)
K_0(i)+2K_2(i)$. Straightforwardly,
\begin{equation}\label{eq:defal3}
\gamma(i) = (n-2i)^2-1.
\end{equation}
Obviously,
$$
\gamma(i) = 0 \qquad \mbox{for } i\in \{(n-1)/2,(n+1)/2\}
$$
and $\gamma(i)>0$ for any other integer $i$.

With an argument similar to that for (\ref{eq:n-3a0+2a2}), we have
$$ (n-1)B_0+2B_2\le \lambda n-2+\sigma $$
(the equality implies $B_0=1$ if $\sigma=1$, 
but for even $\lambda$ we cannot make this conclusion).
Then, we get

\begin{IEEEeqnarray}{rCl}
\label{eq:al3} 2\gamma(0)B'_0 & = & \gamma(0)B'_0 + \gamma(n)B'_n
\leq \sum_i \gamma(i) B'_i\nonumber \\
& = & \frac{2^n((n-1)B_0+2B_2)}{|C|}\\
&  \leq & \frac{2^n(\lambda n-2+\sigma)}{|C|}\nonumber
\end{IEEEeqnarray}
and hence
\[
|C| \leq \frac{2^n(\lambda n-2+\sigma)}{2 \gamma(0)B'_0} =
\frac{2^{n-1}(\lambda n-2+\sigma)}{(n-1)(n+1)}.
\]

(d) Let $n\equiv 0 \bmod 4$. Define  $\delta(i) = n K_0(i)+2K_2(i)$.
Straightforwardly, $ \delta(i) = (n-2i)^2 \ge 0$, and $\delta(i) = 0
\Leftrightarrow i=n/2.$

From the $\lambda$-fold packing condition, we have
$$ nB_0+2B_2\le \lambda n$$
(in this case, the equality does not required $B_0$
to be $1$, so multiple codewords are allowed).
Then, we get
\begin{IEEEeqnarray}{rCl}\nonumber
 2\delta(0)B'_0 & = & \delta(0)B'_0 + \delta(n)B'_n \\ \label{eq:al4} &\leq&
\sum_i \delta(i) B'_i   =   \frac{2^n(nB_0+2B_2)}{|C|}
 \leq \frac{2^n \lambda n}{|C|}
\end{IEEEeqnarray}
and hence
\[
|C| \leq \frac{2^n \lambda n}{2 \delta(0)B'_0} =
\frac{2^{n-1}\lambda}{n}.
\]
\end{IEEEproof}

\begin{corollary}\label{cor:wd}
 Assume that $C$ is an even-weight $\lambda$-fold $1$-packing of length $n$,
 and assume that one of equations (a)--(d) in Theorem~\ref{th:main1}, in respect to $n \bmod 4$,
 is satisfied with equality. Then 
 \begin{itemize}
  \item[\rm (i)] if $n\equiv 1,2 \bmod 4$,
  or $n\equiv 3 \bmod 4$ and $\lambda$ is odd,
  then $C$ is simple
 (there are no multiple codewords);
  \item[\rm (ii)] 
  if $C$ is simple, 
  then
  its weight distribution with respect 
  to any codeword $c\in C$
 is uniquely determined by the parameters $n$ and $\lambda$; in particular, there are exactly
 $\lfloor n(\lambda - 1)/2\rfloor$ codewords at distance $2$ from $c$.
 \end{itemize}
\end{corollary}
\begin{IEEEproof}
 Assume that $C$ is an even-weight $\lambda$-fold $1$-packing of length $n$, $n\equiv 1 \bmod 4$,
 and assume that the inequality (a) in Theorem~\ref{th:main1}  is satisfied with equality.
 This means that we have equalities everywhere in  (\ref{eq:al}).
 As follows from (\ref{eq:n-3a0+2a2}) and the note after it,
 the equality in (\ref{eq:al}) implies $B_0=1$ (which proves claim (i)), $B_2=\lfloor n(\lambda - 1)/2\rfloor$,
 and $B_{n-1}=\lambda$. Since  $A_0(x)\ge 1$, $A_2(x)\le \lfloor n(\lambda - 1)/2\rfloor$,
 and $A_{n-1}(x)\le\lambda$ for every codeword $x$, we also have
 $A_0(x)= 1$, $A_2(x)= \lfloor n(\lambda - 1)/2\rfloor$,
 and $A_{n-1}(x)=\lambda$. Remind also that $B_i=A_i(x)=0$ for every odd $i$.

 Next, consider the dual distance distribution $\{B'_i\}_{i=0}^n$.
 From  (\ref{eq:4cases}) and the equality in (\ref{eq:al}) we find that
 $B'_i=0$ for all $i$ except $0$, $(n-3)/2$, $(n-1)/2$, $(n+1)/2$, $(n+3)/2$, $n$.
 Moreover, we know that $B'_i=B'_{n-i}$ for all $i$ and $B'_0=B'_n=1$.
 So, for complete determining $\{B'_i\}_{i=0}^n$,
 it remains to know $B'_{(n-3)/2}$ and $B'_{(n-1)/2}$.
 These two values can be found from two equations (\ref{eq:B=A}), $k=0,2$.
 So, the dual distance distribution and, hence,
 the distance distribution are uniquely determined.

 The same arguments can be applied to the dual weight distribution 
 $\{A'_i(x)\}_{i=0}^n$
 calculated from $\{A_i(x)\}_{i=0}^n$ by the same formulas as
 $\{B'_i\}_{i=0}^n$ from $\{B_i\}_{i=0}^n$ (\ref{eq:B=A}).
 Indeed, by \cite[Theorem 7(b) in Ch.5, \S 5]{MWS},
 $B'_i=0$ implies $A'_i(x)=0$.
 We also have $A'_i(x)=A'_{n-i}(x)$ and $A'_0(x)=A'_{n}(x)=1$,
 and we know $A_0(x)$ and $A_2(x)$.
 So, we can completely determine $\{A'_i(x)\}_{i=0}^n$ and then $\{A_i(x)\}_{i=0}^n$.

 For $n\equiv 2,3,0 \bmod 4$, the proof is similar.
\end{IEEEproof}

\subsection{Multiple coverings}\label{s:covering}
A set $C$ of vertices of $H(n,q)$  is called a \emph{$q$-ary
$(n,\cdot ,r,\mu)$ multiple covering} \cite[Ch.\,14]{CHLL}
if for every vertex $x$ of
$H(n,q)$ the number of elements of $C$ at distance at most $r$ from
$x$ is not less than $\mu$. 
 Trivially, the complement of any simple $\lambda$-fold $r$-packing 
 in $H(n,q)$
 is a $q$-ary $(n,\cdot ,r,\mu)$ multiple covering, where 
 $\mu = |B_r|-\lambda$, and vice versa. As a consequence, the following lower bounds on the size of binary radius-$1$ coverings are derived from Theorem~\ref{th:main1_}.
 \begin{theorem}
\label{th:main2} The minimum cardinality $K(n,1,\mu)$ of 
a binary $(n,\cdot,1,\mu)$ covering, where  $\mu \equiv \tau \bmod 2$,
$\tau \in \{0,1\}$, satisfy
\begin{enumerate}
 \item[\rm (a)] 
 $\displaystyle K(n,1,\mu) \geq \frac{2^{n}(\mu n+3\mu+\tau)}{n(n+4)}$
 \qquad \mbox{if $n\equiv 0 \bmod 4$}, \\
 \item[\rm (b)] 
 $\displaystyle K(n,1,\mu) \geq  \frac{2^{n}(\mu n+\mu-2)}{(n-1)(n+3)}$
 \qquad \mbox{if $n\equiv 1 \bmod 4$}, \\
 \item[\rm (c)] 
 $\displaystyle K(n,1,\mu) \geq \frac{2^{n}(\mu n+\mu+\tau)}{n(n+2)}$
 \qquad \mbox{if $n\equiv 2 \bmod 4$}, \\
 \item[\rm (d)] 
 $\displaystyle K(n,1,\mu) \geq \frac{2^{n}\mu}{n+1}$
 \qquad \mbox{if $n\equiv 3 \bmod 4$}.
 \end{enumerate}
\end{theorem}
 
 These bounds update the previous lower bounds \cite{HHKL93:cover,Seuranen:2007} in 
 the table \cite[Table~1]{Seuranen:2007} of small values for $K(n,1,\mu)$ in the following positions.
 $$
 \begin{array}{c|c|c|c}
  n & \mu=2 & \mu=3 & \mu=4  \\ \hline
  8 & \mathbf{59}-64  &  \mathbf{91}-94^{\text{\cite{HHKL93:cover}}}  & \mathbf{118}-124^{\text{\cite{Ost:95:cov} }} \\
  10 & {188}^{\text{\cite{Seuranen:2007}}}-216^{\text{\cite{Ost:95:cov}}} & \mathbf{291}-316^{\text{\cite{Ost:95:cov}}} & \mathbf{376}-408^{\text{\cite{Ost:95:cov}}} \\
  12 & \mathbf{640}-704^{\text{\cite{Ost:95:cov}}} & \mathbf{982}-1024 & \mathbf{1280}-1344^{\text{\cite{Ost:95:cov}}} \\
  14 & \mathbf{2195}-2560 & \mathbf{3365}-3712^{\text{\cite{Ost:95:cov}}} & \mathbf{4389}-4864^{\text{\cite{Ost:95:cov}}} \\
  16 & \mathbf{7783}-8192 & \mathbf{11879 }- 12288 & \mathbf{15565 }- 16384 \\ \hline  
 \end{array}
 $$

%==!==!==!==!==!==!==!==!==!==!==!==!==!==!==!==!
%==!==!==!==!==!==!==!==!==!==!==!==!==!==!==!==!
%==!==!==!==!==!==!==!==!==!==!==!==!==!==!==!==!
\section{Optimal packings}\label{s:9}

In this section, we discuss some optimal packings,
namely,  binary two-fold packings attaining 
the bound in Theorem~\ref{th:main1_} for small $n$
and multifold $1$-packings attaining the sphere-packing bound.
Many of the discussed packings are related with very regular objects called equitable partitions. 

A partition $\pi = (C_0, C_1,  \ldots , C_{m})$ of the vertices of a
graph (in our case, $H(10,2)$) is called \emph{equitable} if for every
$i$ and $j$ from $\{0, 1, \ldots, m \}$ there is an integer $s_{i,j}$
such that each vertex $v$ in $C_i$ has exactly $s_{i,j}$ neighbors
in $C_j$. The matrix $(s_{i,j})_{i,j=0}^{m}$ is called the
\emph{intersection matrix} (sometimes, the quotient matrix of the
graph with respect to the partition). If it is tridiagonal
(equivalently, $\pi$ is a distance partition with respect to $C_0$),
then $C_0$ is called a \emph{completely regular code} 
of \emph{covering radius} $m$ with
{intersection matrix} $(s_{i,j})_{i,j=0}^{m}$ 
(alternatively, the parameters of a completely regular code are often given in the form of the \emph{intersection array} 
$(s_{0,1},\ldots,s_{m-1,m};s_{1,0},\ldots,s_{m,m-1})$), see e.g. the survey \cite{BRZ:CR} for the background of this important concept.

%==!==!==!==!==!==!==!==!==!==!==!==!==!==!==!==!
\subsection{Optimal $2$-fold packings in $H(n,2)$ for small $n$}\label{s:2-9}

As special cases of Theorems~\ref{th:main1_} and~\ref{th:main1}, we have 
the bounds for $\PPP 22n1$ reflected in Table~\ref{t:9}.
\begin{table}
$$
\begin{array}{l||cc|cc|cc|cc}
n         & 2 & 3 & 4 & 5  & 6  &  7 &  8 & 9  \\ \hline
\PPP n221 & 2 & 4 & 5 & 10 & 16 & 32 & 48 & 96 \\
N(n)      & 2 & 1 & 1 & 1  & 7  &  3 & 20 & 6  \\
N^*(n+1)  & 1 & 1 & 1 & 1  & 3  &  3 &  6 & 3  \\
\end{array}
$$
 \caption{Numbers $N(n)$ and $N^*(n+1)$ of inequivalent
 $2$-fold $1$-packings in $H(n,2)$ and even-weight $2$-fold $1$-packings in $H(n+1,2)$, respectively, $n\le 9$. }\label{t:9}
\end{table}
As a result of the exhaustive computer-aided 
search (see the appendix for the description of the algorithm), we found that all these bounds for $n\le 9$ are tight,
and we found the number of all equivalence classes of simple optimal $2$-fold 
$1$-packings in $H(n,2)$, $n\le 9$, and of their extensions in $H(n+1,2)$.
It happens that each of packing with odd $n\le 9$ from the classification,
as well as each even-weight packing with any $n\le 10$, 
can be represented as a cell of an equitable partition with special intersection
matrix, depending on the parameters of the packing:

% There are exactly $3$ equivalence
% classes of even-weight $2$-fold $1$-packings of size $96$ in
% $H(10,2)$ and $6$ equivalence
% classes of even-weight $2$-fold $1$-packings of size $48$ in
% $H(9,2)$. By projecting, we get $6$ equivalence
% classes of $2$-fold $1$-packings of size $96$ in
% $H(9,2)$ and $20$ equivalence
% classes of $2$-fold $1$-packings of size $48$ in
% $H(8,2)$. 
% Each of the considered optimal packings
% in $H(10,2)$ and $H(9,2)$
% (but not in $H(8,2)$)
% can be represented as the last cell an equitable partition 
% % $\pi=(C_0,C_1,C_2,C_3,C_4)$
% with intersection matrix 
% $S^{10}_{\mathrm{even}}$, 
% $S^{9}_{\mathrm{even}}$,
% or $S^{9}$, respectively,

$$
S^{n+1}_{\mathrm{even}}=
 \left(\begin{matrix} 
  0 & n+1 &  0  &  0 &  0 \cr 
  1 &  0  &  n  &  0 &  0 \cr 
  0 & n-3 &  0  &  2 &  2 \cr 
  0 &  0  & n+1 &  0 &  0 \cr
  0 &  0  & n+1 &  0 &  0 
 \end{matrix} \right),
\quad
S^{n}_{\mathrm{even}}=
\left(
\begin{matrix}
 0_{4\times 4}    & S^{n} \\
 S^{n} & 0_{4\times 4}      
\end{matrix}
\right), \quad
 S^{n}=
 \left(\begin{matrix}
 0 & n   & 0 & 0 \cr
 1 & n-5 & 2 & 2 \cr
 0 & n-3 & 1 & 2 \cr
 0 & n-3 & 2 & 1 
 \end{matrix} \right), 
 \quad
 n=5,9;
$$ 
$$
S^{n+1}_{\mathrm{even}}=
 \left(\begin{matrix} 
  0   &  n-1 &  2   \cr 
  n+1 &  0   &  0   \cr 
  n+1 &  0   &  0   
 \end{matrix} \right),
\qquad
S^{n}_{\mathrm{even}}=
\left(
\begin{matrix}
 0_{2\times 2}    & S^{n} \\
 S^{n} & 0_{2\times 2}      
\end{matrix}
\right), \qquad
 S^{n}=
 \left(\begin{matrix}
 n-2 & 2 \cr
 n-1 & 1  
 \end{matrix} \right), 
 \qquad n=3,7.
$$
% $$
%  S^{9}_{\mathrm{even}}=
% \text{
% $\small
%  \left(\begin{array}{cccc|cccc} 
%  0 & 0 & 0 & 0  &  0 & 9 & 0 & 0 \cr
%  0 & 0 & 0 & 0  &  1 & 4 & 2 & 2 \cr
%  0 & 0 & 0 & 0  &  0 & 6 & 1 & 2 \cr
%  0 & 0 & 0 & 0  &  0 & 6 & 2 & 1 \cr\hline
%  0 & 9 & 0 & 0  &  0 & 0 & 0 & 0 \cr
%  1 & 4 & 2 & 2  &  0 & 0 & 0 & 0 \cr
%  0 & 6 & 1 & 2  &  0 & 0 & 0 & 0 \cr
%  0 & 6 & 2 & 1  &  0 & 0 & 0 & 0 
%  \end{array} \right). 
%  $
%  }
% $$
It is not difficult to see the inverse: 
the last cell of an equitable partition with one of the mentioned 
intersection matrices is an optimal (even/odd-weight) two-fold $1$-packing 
(the proportions between the cardinalities of the cells 
can be seen from the matrix, 
as well as the parity of each cell in the case of 
$S^{...}_{\mathrm{even}}$;
the packing condition can be seen from the last column).

We refer the connection with equitable partitions as 
a computational result.
However, for some parameters 
% (including the ones related with the matrix 
% $S^{10}_{\mathrm{even}}$) 
it can be explained 
theoretically with the technique developed in 
\cite{Kro:2m-3,Kro:2m-4}, may be generalizing
to packings attaining the bound
with some other special values of $n$ and $\lambda$ 
(in \cite{Kro:2m-3,Kro:2m-4} it was shown 
for $\lambda=1$, $n=2^m-2$ and  $n=2^m-3$,
respectively,
with corresponding intersection matrices).
Here, we do not focus on the details of that theory,
because 
% reproducing the proof would not cover all the cases 
% and would not bring any thing new to the theory
the number of cases we are interested is finite,
and the proof from \cite{Kro:2m-3,Kro:2m-4} works
without essential changes for all considered parameters
except the ones related with the matrix 
$S^{9}_{\mathrm{even}}$. 

Another empiric fact is that every 
$2$-fold $1$-packing in $H(n,2)$,
$n=2,4,6,8$,
can always be uniquely represented as 
a shortened $2$-fold $1$-packing of length $n+1$
(similarly, 
for even-weight extensions). Again, we have theoretical 
explanations for some parameters, but not for all.
Similar natural questions were previously 
considered in the case
$\lambda=1$: every optimal $1$-packing in $(n,2)$,
$n=2^m-2$, is always a shortened 
optimal $1$-packing in $(n+1,2)$ \cite{Bla99},
while this is not always true for $n=2^m-3$ and $n=2^m-4$
\cite{OstPot:13-512-3,KOP:2011}.

The most interesting $2$-fold $1$-packings 
from the considered classification up to length $9$
are the optimal packings
in $H(9,2)$ and their length-$10$ extensions
(indeed, the packings in $H(8,2)$ 
are obtained by shortening 
from the packings in $H(9,2)$; 
the size of the packings in $H(7,2)$ and $H(6,2)$ 
is not better than the union of two cosets of 
a $1$-error-correcting code; 
the extention of the best packing in $H(6,2)$ 
is equivalent to a unique $2$-$(6,3,2)$-design; 
$H(5,2)$ is again a shortened case; 
the other are trivial).
The cardinality $96$ of these packings
is very close to the sphere-packing bound
$\lfloor 2\cdot 2^9 / (1+9)\rfloor = 102$. To
compare, the largest $1$-error-correcting code 
in $H(9,2)$ has $40$
codewords \cite{Best80}, 
and the union of two disjoint such codes is
a $2$-fold $1$-packing of cardinality $80$ only.
The intersection matrix 
$S_{\mathrm{even}}^{10}$
of the equitable
partition corresponding to the extended packings 
is not tridiagonal, 
but unifying the last two cells 
(which are in fact an even-weight 
$2$-fold packing $P$ of size $96$ 
and its antipode $P+1111111111$)
results in an equitable partition with 
tridiagonal intersection matrix. 
So, the first cell of the equitable partition is
a completely regular code with intersection array
$(10,9,4;1,6,10)$. 
One such code, the only linear one, 
was already known 
\cite[Theorem~1(2)]{RifZin:2009};
some details about all three nonequivalent
completely regular codes,
including their propelinear structures, 
and corresponding
optimal $2$-fold packings can be found
in the proceedings~\cite{KroPot:ISIT2019:2fold} 
(those details were considered to be too peculiar to include them in the current paper). 
Here, we present only one example of 
an optimal even-weight $2$-fold $1$-packing 
of size $96$:
\begin{IEEEeqnarray*}{l}
 \langle 0001111011,\ 0010101010,\ 0100110100,\ 1000110111\rangle \\
 \ \ \ {} +
 \{0000000000,0000100001,0000100111, \\ \ \ \ \ \ \ \ \ \ \  0000101110,0000111001,0000111100\}.
\end{IEEEeqnarray*}

% Unifying the cells $C_3$ and $C_4$, 
% we obtain an equitable partition
% with tridiagonal intersection matrix.
% % \begin{equation}\label{eq:C0123}
% % \left(\matrix{0&10&0&0 \cr 1&0&9&0 \cr 0&6&0&4 \cr 0&0&10&0} \right).
% % \end{equation}
% So, $C_0$ is a completely regular code. One such code was already
% known \cite[Theorem~1(2)]{RifZin:2009}; it is linear of dimension
% $5$ and equivalent to $C_0$ when $C_4$ is the unitrade
% \begin{IEEEeqnarray*}{l}
%  \langle 0001111011,\ 0010101010,\ 0100110100,\ 1000110111\rangle \\
%  \ \ \ {} +
%  \{0000000000,0000100001,0000100111, \\ \ \ \ \ \ \ \ \ \ \  0000101110,0000111001,0000111100\}
% \end{IEEEeqnarray*}
%  from our classification. 
%  The other two unitrades of cardinality $96$ correspond to nonlinear
% completely regular codes.
% These two codes however also have nice propelinear structure, in the sense of \cite{RifPuj:1997},
% and one of them is $Z_2Z_4$-linear (see, e.g., \cite{BFPRV2010} for the background of  $Z_2Z_4$-linear codes).
% The detailed description of these codes and corresponding unitrades can be found in 
% \cite{KroPot:ISIT2019:2fold} 
% (those details were considered to be too peculiar to include them in the current paper).

\begin{remark}
  We note that our classification results do not imply 
 the
 nonexistence of other length-$10$ completely regular codes with 
 intersection array $(10,9,4;1,6,10)$. 
 However, this can be easily derived from the following known 
 facts: (i) if such code contains the all-zero word,
 then there are $15$ weight-$4$ codewords 
 and they form a $2$-design, see \cite[Th.\,8]{GoeTil:UPC};
 (ii) any completely regular code $C$ with considered parameters
 is self-complementary, that is, $C=C+1111111111$, see, e.g., \cite{BRZ:2008}; 
 (iii) there are exactly $3$ isomorphism classes of $2$-$(10,4,2)$ designs,
 see \cite[Table~1.25]{MatRos:small}.
 So, by some ``magic'' reason, the set of vertices at maximum distance from each
 binary length-$10$ completely regular code with 
 intersection array $(10,9,4;1,6,10)$ can be splitted into two $2$-fold $1$-packings,
 in more than one way,
 and all ways to splits result in equivalent packings, for the same code.
\end{remark}

%==!==!==!==!==!==!==!==!==!==!==!==!==!==!==!==!
\subsection{Multifold perfect codes}\label{s:perfect}

By analogy with classical perfect codes, attaining the sphere-packing
bound, a $\lambda$-fold $r$-packings of cardinality
$\lambda q^n/|B_r|$ (see \eqref{eq:spb}) in $H(n,q)$ can be called \emph{perfect}, or alternatively, 
a \emph{$\lambda$-fold $r$-perfect code}.
Moreover, such packings are at the same time optimal $\lambda$-fold coverings, and they also known as \emph{perfect multiple coverings}~\cite[Section~14.2]{CHLL}. 
The existence of multiple coverings with radius larger than $1$
is a complicate problem, especially for small $\lambda$.
A survey can be found in \cite[Section~14.2]{CHLL}; 
in \cite{Vorobev:2012:multiple}, 
Vorob'ev considered 
multifold $r$-perfect codes in $H(n,2)$, $r>1$,
related with equitable $2$-partitions.
In the rest of this section,
we focus on the case $r=1$.
For a $\lambda$-fold $1$-pefrect code $C$, 
every radius-$1$ ball in $H(n,q)$ contains exactly 
$\lambda$ codewords. Moreover, if $r=1$ and $C$ has no multiple codewords, then $C$ and its complement form an equitable partition with intersection matrix
$$
\left(
\begin{array}{cc}
 \lambda-1 & (q-1)n-\lambda+1 \\
 \lambda & (q-1)n-\lambda \\
\end{array}
\right).
$$
The famous Lloyd condition (see, e.g., \cite[Theorem~9.3.3]{GoRo} for equitable partitions) says that the eigenvalue $-1$ of this intersection matrix must belong to the eigenspectrum
$\{-n+iq \mid i\in\{0,\ldots,n\}\}$ of the Hamming graph $H(n,q)$,
which is equivalent to $n\equiv 1\bmod q$
(for example, this shows that a $5$-fold $1$-perfect code of size $32$ in $H(3,4)$ does not exist).
Existence of such partitions in the binary case 
was shown in~\cite{FDF:PerfCol}, as a part of a general theory
of equitable $2$-partitions of $H(n,2)$ 
(which was partially generalized to an arbitrary $q$ in a recent work~\cite{BKMTV}), but the sufficiency of the sphere-packing 
condition for the existence of multifold $1$-perfect codes 
was known earlier for any prime $q$:
\begin{proposition}[{\cite[Theorem~14.2.4]{CHLL}}]\label{th:perfect}
Assume that $q$ is prime. Simple (without multiple codewords) $\lambda$-fold 
$1$-perfect codes in $H(n,q)$
exist if and only if $K=\lambda q^n/|B_1|$ is integer and $\lambda\le |B_1|$, where
$B_1$ is a radius-$1$ ball in $H(n,q)$, $|B_1|=(q-1)n+1$.
\end{proposition}

The sufficiency in Proposition~\ref{th:perfect} is proved by
constructing a code as the union of $\kappa$ cosets of a linear  $\lambda/\kappa$-fold $1$-perfect code of dimension $k$
(defined, for example, by a parity-check matrix,
generalizing the Hamming code (see, e.g., \cite[p.193]{MWS})), 
where
\begin{equation}\label{eq:kK}
 K=\frac{\lambda q^n}{(q-1)n+1}=\kappa q^k, \qquad \gcd(\kappa,q)=1.
\end{equation}
The same argument works to construct 
$\lambda$-fold $1$-perfect codes in $H(n,q)$ for any 
prime-power $q$; so, \eqref{eq:kK} is sufficient
for the existence, but not necessary: if $q$ is not prime,
then not all integer $K$ are representable in the form \eqref{eq:kK}. The smallest example (also satisfying the Lloyd condition) is $q=4$, $n=13$, $\lambda=5$. In this case,
$K=2^{23}=2\cdot 4^{11}$, 
and we see that $\lambda/\kappa$ is not integer
for any representetion of $K$ in the form $\kappa 4^k$.
So, a code cannot be constructed as the union of cosets of a linear multifold $1$-perfect code.
However, we can construct a $5$-fold $1$-perfect code as an additive code: we first construct a binary linear code 
as the kernel of the check matrix 
$$
H=\left(
\begin{array}{c@{\,}c@{~}c@{\,}c@{~}c@{\,}c@{~}c@{\,}c@{~}c@{\,}c@{~}c@{\,}c@{~}c@{\,}c@{~}c@{\,}c@{~}c@{\,}c@{~}c@{\,}c@{~}c@{\,}c@{~}c@{\,}c@{~}c@{\,}c}
0 & 0  &  0 & 0  &  0 & 1  &  0 & 1  &  1 & 0  &  1 & 0  &  0 & 1  &  0 & 1  &  1 & 0  &  0 & 1  &  0 & 1  &    1 & 0  &   0 & 0  \cr
0 & 1  &  0 & 1  &  1 & 0  &  1 & 0  &  0 & 0  &  0 & 0  &  0 & 1  &  1 & 0  &  0 & 1  &  1 & 1  &  1 & 1  &    1 & 0  &   0 & 0  \cr
1 & 0  &  1 & 0  &  0 & 0  &  0 & 0  &  0 & 1  &  0 & 1  &  1 & 0  &  0 & 1  &  0 & 1  &  1 & 0  &  1 & 0  &    1 & 0  &   0 & 0  
\end{array}
\right),
$$
then treat binary words of length $26$ as quaternary words of length $13$ in the alphabet $\{00$, $01$, $10$, $11\}$.
To make sure that the code is $5$-fold $1$-perfect, one can check that each of $8$ different syndromes (height-$3$ binary columns)
is represented in exactly $5$ ways as $H e^{\mathrm{T}}$ where $e$ is one of $40$ words of weight at most $1$ (in the metric of $H(13,4)$):
$00 ... 00$, 
$\underline{01} 00 ... 00$, 
$\underline{10} 00 ... 00$,  
$\underline{11} 00 ... 00$, 
$00 \underline{01} 00 ... 00$, \ldots, 
$00 ... 00\underline{11}$.

\begin{problem}
Is the integrality of $\frac{\lambda q^n}{(q-1)n+1}$
sufficient for the existence of 
$\lambda$-fold $1$-perfect codes in $H(n,q)$,
where $q$ is a prime power but not prime?
\end{problem}

If $q$ is not a prime power, then the problem of the existence 
of nontrivial multifold $1$-perfect codes is open, generalizing
the famous problem of the existence of perfect codes. Up to our
knowledge, no notrivial multifold $1$-perfect codes are known 
over a none-prime-power alphabet.

% Much less is known about the multifold $r$-perfect codes with $r>1$. 
% In~\cite{Vorobev:2012:multiple}, Vorob'ev considered 
% multifold $r$-perfect codes in $H(n,2)$ that are at the same time cells
% of equitable $2$-partitions and found some nontrivial examples,
% for example, a $7$-fold $2$-perfect code in $H(10,2)$.

%We finalise this section by noting that 
In the binary case,
$\lambda$-fold $1$-perfect codes attain the bound of Theorem~\ref{th:main1_}(d), which trivially coincides with the sphere-packing bound; moreover, after shortening, 
$2$-fold $1$-perfect codes turn to packings attaining the bound of Theorem~\ref{th:main1_}(c). This observation exhaust the known infinite
series of $\lambda$-fold $1$-packings attaining the bounds of Theorem~\ref{th:main1_} (Theorem~\ref{th:main1}, for extended packings) with
$1<\lambda\le n$. 
Moreover, together with the results of Section~\ref{s:2-9},
it cover all possible $2$-fold packings lying on those bounds 
(it is not difficult to find that
$\frac{2^n\cdot 2n}{(n-1)(n+3)}$ can be integer only if $n$ is 
$5$ or $9$).
In the case $\lambda=1$, 
the bounds turn to the bounds
in~\cite{BesBro77} and are attained by $t$-times-shortened 
$1$-perfect codes, $t=0,1,2,3$.

\begin{problem}
 Find more examples of $\lambda$-fold $1$-packings attaining the bound
 in Theorem~\ref{th:main1_}. 
 In particular, does there exist a $3$-fold $1$-packing of length $8$ and cardinality $80$,
 lying on bound~(a)?
\end{problem}

%==!==!==!==!==!==!==!==!==!==!==!==!==!==!==!==!
%==!==!==!==!==!==!==!==!==!==!==!==!==!==!==!==!
%==!==!==!==!==!==!==!==!==!==!==!==!==!==!==!==!
\section{$1$-perfect unitrades and extended $1$-perfect unitrades}\label{s:uni}

In this section, we consider the so-called $1$-perfect unitrades, 
which are a special case of two-fold $1$-packing.
The term ``unitrade'' was introduced in \cite{Potapov:2013:trade},
where latin bitrades (see Remark~\ref{r:latin}) 
were studied and some of their properties were described in terms of unitrades.
Here we consider another kind of unitrades; 
however, as one can see from the sequence of remarks in this section,
the class of $1$-perfect unitrades is also connected with latin bitrades
(see e.g. \cite{Cav:rev}, \cite{Potapov:2013:trade})
and trades of combinatorial designs (see e.g. \cite{HedKho:trades}),
well known in the corresponding areas of combinatorics.

A \emph{$1$-perfect unitrade} is a set $T$ of vertices of $H(n,q)$ such that $ |B\cap T|
\in \{0,2\}$ for every radius-$1$ ball $B$ in $H(n,q)$.
In the case $q=2$, appending the parity check bit turns a
$1$-perfect unitrade in $H(n,2)$ to a so-called 
extended $1$-perfect unitrade in $H(n+1,2)$, 
defined as follows.
 A set $T$ of even-weight (or odd-weight) vertices of $H(m,2)$  such that $ |B\cap T|
\in \{0,2\}$ for every radius-$1$ ball $B$ in $H(m,2)$ 
centered in an odd-weight (respectively, even-weight) word
 is called an \emph{extended $1$-perfect unitrade}.
 Clearly, removing the last coordinate in all words of an 
extended $1$-perfect unitrade in $H(n+1,2)$ turns it 
to a $1$-perfect unitrade in $H(n,2)$; so, the $1$-perfect unitrades
are in one-to-one correspondence with the even-weight (or odd-weight)
extended $1$-perfect unitrades.

\begin{proposition}\label{p:parity}
 (i) If $q$ is even, then non-empty 
 $1$-perfect unitrades in $H(n,q)$ can only exist
 if $n$ is odd. 
 (ii) Nonempty extended $1$-perfect unitrades in $H(n,2)$ 
 can only exist if $n$ is even.
\end{proposition}
\begin{IEEEproof}
We claim that if $x$ is a word from a 
 $1$-perfect unitrade $U$ in $H(n,q)$, 
 then $U$ has exactly one word at distance $1$ from $x$ and exactly $(n-1)(q-1)/2$ words at distance $2$
 from $x$.
 Indeed, at first, by the unitrade definition, a radius-$1$ ball centered in $x$ has exactly one another word in $U$, say $y$.
 At second, each of $(n-1)(q-1)$ neighbors of $x$ 
 that are not neighbors of $y$ has another neighbor in $U$, say $z$,
 and $z$ is at distance $2$ from $x$. On the other hand, each vertex 
 at distance $2$ from $x$ has exactly $2$ common neighbors with $x$; 
 hence, the number of such vertices in $U$ is $(n-1)(q-1)/2$.
 Since this value is integer, (i) is proven; (ii) is straightforward from (i).
\end{IEEEproof}

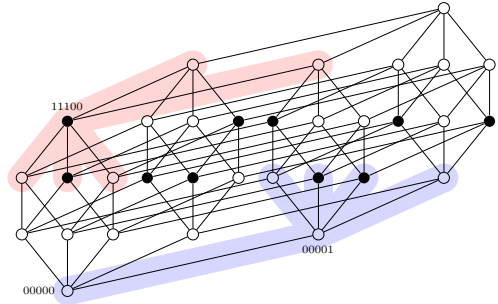
\begin{figure}[bt]
 \centering
 \scalebox{0.6}
 {
 \begin{tikzpicture}[scale=2.5]
 \def\aX{-0.4} \def\aY{+0.5}
 \def\bX{+0.0} \def\bY{+0.5}
 \def\cX{+0.4} \def\cY{+0.5}
 \def\dX{+1.1} \def\dY{+0.5}
 \def\eX{+2.2} \def\eY{+0.5}
 \def\bllsZ{18pt}
 \def\bllsz{18pt}
 \tikzstyle{zr}=[circle,minimum size=0pt,inner sep=0pt]
 \tikzstyle{vtX}=[circle,minimum size=7pt,inner sep=0pt]
 \tikzstyle{bll}=[circle,minimum size=\bllsZ,inner sep=0pt, fill=red!16]
 \tikzstyle{bla}=[circle,minimum size=\bllsZ,inner sep=0pt, fill=blue!16]
 \tikzstyle{vtx} = [vtX, thin, draw=black]
 \tikzstyle{vtC} = [vtX, fill=black]
 \tikzstyle{selected edge} = [draw,line width=5pt,-,red!50]
 \tikzstyle{edge} = [draw,-,black]
 \node[bll] (b00000) at ( \cX\bX\aX      , \cY\bY\aY       ) {}; % {00111};
 \node[bll] (b10000) at ( \eX\cX\bX\aX   , \eY\cY\bY\aY    ) {}; % {10111};
 \node[bll] (b01000) at ( \dX\cX\bX\aX   , \dY\cY\bY\aY    ) {}; % {01111};
 \node[bll] (b00100) at ( \bX\aX         , \bY\aY          ) {}; % {00011};
 \node[bll] (b00010) at ( \cX\aX         , \cY\aY          ) {}; % {00101};
 \node[bll] (b00001) at ( \cX\bX         , \cY\bY          ) {}; % {00110}; 
 \node[zr] (B00000) at ( \cX\bX\aX      , \cY\bY\aY       ) {}; % {00111};
 \node[zr] (B10000) at ( \eX\cX\bX\aX   , \eY\cY\bY\aY    ) {}; % {10111};
 \node[zr] (B01000) at ( \dX\cX\bX\aX   , \dY\cY\bY\aY    ) {}; % {01111};
 \node[zr] (B00100) at ( \bX\aX         , \bY\aY          ) {}; % {00011};
 \node[zr] (B00010) at ( \cX\aX         , \cY\aY          ) {}; % {00101};
 \node[zr] (B00001) at ( \cX\bX         , \cY\bY          ) {}; % {00110};
 \draw[draw,line width=\bllsz,-,red!16] (B10000) -- (B00000) -- (B01000);
 \draw[draw,line width=\bllsz,-,red!16] (B00100) -- (B00000) -- (B00010);
 \draw[draw,line width=\bllsz,-,red!16] (B00000) -- (B00001);
 \node[bla] (a00000) at ( \eX            , \eY             ) {}; % {10000};
 \node[bla] (a10000) at ( 0              , 0               ) {}; % {00000};
 \node[bla] (a01000) at ( \eX\dX         , \eY\dY          ) {}; % {11000};
 \node[bla] (a00100) at ( \eX\cX         , \eY\cY          ) {}; % {10100};
 \node[bla] (a00010) at ( \eX\bX         , \eY\bY          ) {}; % {10010};
 \node[bla] (a00001) at ( \eX\aX         , \eY\aY          ) {}; % {10001};
 \node[zr] (A00000) at ( \eX            , \eY             ) {}; % {10000};
 \node[zr] (A10000) at ( 0              , 0               ) {}; % {00000};
 \node[zr] (A01000) at ( \eX\dX         , \eY\dY          ) {}; % {11000};
 \node[zr] (A00100) at ( \eX\cX         , \eY\cY          ) {}; % {10100};
 \node[zr] (A00010) at ( \eX\bX         , \eY\bY          ) {}; % {10010};
 \node[zr] (A00001) at ( \eX\aX         , \eY\aY          ) {}; % {10001};
 \draw[draw,line width=\bllsz,-,blue!16] (A10000) -- (A00000) -- (A01000);
 \draw[draw,line width=\bllsz,-,blue!16] (A00100) -- (A00000) -- (A00010); 
 \draw[draw,line width=\bllsz,-,blue!16] (A00000) -- (A00001); 
 \node[vtx,label=left:$_{00000}$] (v00000) at ( 0              , 0               ) {}; % {00000};
 \node[vtx] (v00001) at ( \aX            , \aY             ) {}; % {00001};
 \node[vtx] (v00010) at ( \bX            , \bY             ) {}; % {00010};
 \node[vtx] (v00011) at ( \bX\aX         , \bY\aY          ) {}; % {00011};
 \node[vtx] (v00100) at ( \cX            , \cY             ) {}; % {00100};
 \node[vtC] (v00101) at ( \cX\aX         , \cY\aY          ) {}; % {00101};
 \node[vtx] (v00110) at ( \cX\bX         , \cY\bY          ) {}; % {00110};
 \node[vtC,label=above:$_{11100}$] (v00111) at ( \cX\bX\aX      , \cY\bY\aY       ) {}; % {00111};
 \node[vtx] (v01000) at ( \dX            , \dY             ) {}; % {01000};
 \node[vtC] (v01001) at ( \dX\aX         , \dY\aY          ) {}; % {01001};
 \node[vtC] (v01010) at ( \dX\bX         , \dY\bY          ) {}; % {01010};
 \node[vtx] (v01011) at ( \dX\bX\aX      , \dY\bY\aY       ) {}; % {01011};
 \node[vtx] (v01100) at ( \dX\cX         , \dY\cY          ) {}; % {01100};
 \node[vtx] (v01101) at ( \dX\cX\aX      , \dY\cY\aY       ) {}; % {01101};
 \node[vtC] (v01110) at ( \dX\cX\bX      , \dY\cY\bY       ) {}; % {01110};
 \node[vtx] (v01111) at ( \dX\cX\bX\aX   , \dY\cY\bY\aY    ) {}; % {01111};
 \node[vtx,label=below:$_{00001}$] (v10000) at ( \eX            , \eY             ) {}; % {10000};
 \node[vtx] (v10001) at ( \eX\aX         , \eY\aY          ) {}; % {10001};
 \node[vtC] (v10010) at ( \eX\bX         , \eY\bY          ) {}; % {10010};
 \node[vtC] (v10011) at ( \eX\bX\aX      , \eY\bY\aY       ) {}; % {10011};
 \node[vtC] (v10100) at ( \eX\cX         , \eY\cY          ) {}; % {10100};
 \node[vtx] (v10101) at ( \eX\cX\aX      , \eY\cY\aY       ) {}; % {10101};
 \node[vtx] (v10110) at ( \eX\cX\bX      , \eY\cY\bY       ) {}; % {10110};
 \node[vtx] (v10111) at ( \eX\cX\bX\aX   , \eY\cY\bY\aY    ) {}; % {10111};
 \node[vtx] (v11000) at ( \eX\dX         , \eY\dY          ) {}; % {11000};
 \node[vtC] (v11001) at ( \eX\dX\aX      , \eY\dY\aY       ) {}; % {11001};
 \node[vtx] (v11010) at ( \eX\dX\bX      , \eY\dY\bY       ) {}; % {11010};
 \node[vtx] (v11011) at ( \eX\dX\bX\aX   , \eY\dY\bY\aY    ) {}; % {11011};
 \node[vtC] (v11100) at ( \eX\dX\cX      , \eY\dY\cY       ) {}; % {11100};
 \node[vtx] (v11101) at ( \eX\dX\cX\aX   , \eY\dY\cY\aY    ) {}; % {11101};
 \node[vtx] (v11110) at ( \eX\dX\cX\bX   , \eY\dY\cY\bY    ) {}; % {11110};
 \node[vtx] (v11111) at ( \eX\dX\cX\bX\aX, \eY\dY\cY\bY\aY ) {}; % {11111};
 \draw[edge] (v00000) -- (v00010) -- (v00011) -- (v00001) -- (v00000);
 \draw[edge] (v00100) -- (v00110) -- (v00111) -- (v00101) -- (v00100);
 \draw[edge] (v01000) -- (v01010) -- (v01011) -- (v01001) -- (v01000);
 \draw[edge] (v01100) -- (v01110) -- (v01111) -- (v01101) -- (v01100);
 \draw[edge] (v10000) -- (v10010) -- (v10011) -- (v10001) -- (v10000);
 \draw[edge] (v10100) -- (v10110) -- (v10111) -- (v10101) -- (v10100);
 \draw[edge] (v11000) -- (v11010) -- (v11011) -- (v11001) -- (v11000);
 \draw[edge] (v11100) -- (v11110) -- (v11111) -- (v11101) -- (v11100);
 \draw[edge] (v00000) -- (v01000) -- (v01100) -- (v00100) -- (v00000);
 \draw[edge] (v00001) -- (v01001) -- (v01101) -- (v00101) -- (v00001);
 \draw[edge] (v00010) -- (v01010) -- (v01110) -- (v00110) -- (v00010);
 \draw[edge] (v00011) -- (v01011) -- (v01111) -- (v00111) -- (v00011);
 \draw[edge] (v10000) -- (v11000) -- (v11100) -- (v10100) -- (v10000);
 \draw[edge] (v10001) -- (v11001) -- (v11101) -- (v10101) -- (v10001);
 \draw[edge] (v10010) -- (v11010) -- (v11110) -- (v10110) -- (v10010);
 \draw[edge] (v10011) -- (v11011) -- (v11111) -- (v10111) -- (v10011);
 \draw[edge] (v00000) -- (v10000); \draw[edge] (v01000) -- (v11000); 
 \draw[edge] (v00100) -- (v10100); \draw[edge] (v01100) -- (v11100); 
 \draw[edge] (v00010) -- (v10010); \draw[edge] (v01010) -- (v11010); 
 \draw[edge] (v00110) -- (v10110); \draw[edge] (v01110) -- (v11110); 
 \draw[edge] (v00001) -- (v10001); \draw[edge] (v01001) -- (v11001); 
 \draw[edge] (v00101) -- (v10101); \draw[edge] (v01101) -- (v11101); 
 \draw[edge] (v00011) -- (v10011); \draw[edge] (v01011) -- (v11011); 
 \draw[edge] (v00111) -- (v10111); \draw[edge] (v01111) -- (v11111); 
 \end{tikzpicture}
 }
 \caption{A $1$-perfect unitrade, an optimal $2$-fold $1$-packing in $H(5,2)$}
 \label{f:H52}
 \end{figure}

\begin{example}\label{ex:5}
 Consider the following set of binary words:
 $$C=\{ 00101,01010,10100,01001,10010,
    00111,01110,11100,11001,10011\} $$
    (see Fig.~\ref{f:H52}).
    It is a $1$-perfect unitrade in $H(5,2)$. 
    For example, the word $00000\not\in C$ has no neighbors in $C$,
    the word $00001\not\in C$ has two neighbors  $00101$ and $01001$ in $C$,
    % the word $00011\not\in C$ has two neighbors  $00111$ and $10011$ in $C$,
    the word $00111\in C$ has one neighbor $00101$ in $C$ (so, the radius-$1$ ball centered in $00111$ totally has two elements of $C$). 
    Moreover, $C$ is a $2$-fold $1$-packing attaining the bound
    of Theorem~\ref{th:main1_}(b).
    Appending the first $5$ words of $C$ by $1$ and the last $5$, by $0$, we get an extended $1$-perfect unitrade.
\end{example}

\begin{remark}\label{r:halved}
Alternatively, the extended $1$-perfect unitrades can be defined
in terms of cliques of the halved $n$-cube.
A maximum collection of mutually adjacent vertices of a graph 
is called a \emph{maximum clique}. 
%It is easy to see that 
If $n\ge 5$, then 
every maximum clique of $\HH{n}$ consists of $n$ vertices at Hamming distance one
from some odd-weight binary word of length $n$ 
(such a word, of course, is not itself a vertex of $\HH{n}$). 
So, a set $T$ of vertices of $\HH{n}$ 
 is  an {extended $1$-perfect unitrade} if and only if 
$ |B\cap T| \in \{0,2\}$ for every maximum clique $B$ in $\HH{n}$.
\end{remark}

A (extended) $1$-perfect unitrade is called \emph{primary} if it cannot be partitioned into two proper subsets that are (extended) $1$-perfect unitrades too.
A (extended) $1$-perfect trade is called \emph{bipartite} if it can be partitioned into two codes
with minimum distance larger than $2$.

The next three remarks show connections of $1$-perfect unitrades with 
other areas of discrete mathematics such that design theory and latin squares. 
The concepts introduced there are not used in the rest of the paper.
 
\begin{remark}\label{r:perfect}
The $1$-perfect unitrades
are related with  the $1$-perfect bitrades and $1$-perfect codes in the following manner. A \emph{$1$-perfect bitrade} is a pair $(T_+,T_-)$ of disjoint sets of vertices of $H(n,q)$ such that $|B\cap T_+|=|B\cap T_-|\in \{0,1\}$ for every radius-$1$ ball $B$. An \emph{$r$-perfect code} is a set $C$ of vertices of $H(n,q)$ such that $|B\cap C|=1$ for every radius-$r$ ball $B$. 
If $(T_+,T_-)$ is a $1$-perfect bitrade, 
then $T_+ \cup T_-$ is a $1$-perfect unitrade
(but only bipartite $1$-perfect unitrades are representable in such a way).
If $C$ and $C'$ are $1$-perfect codes, 
then $(C\backslash C',C'\backslash C)$ 
is a $1$-perfect bitrade
(but not every $1$-perfect bitrade is representable in such a way).
In a similar manner, extended $1$-perfect unitrades
are related to  the \emph{extended $1$-perfect bitrades} 
and the \emph{extended $1$-perfect binary codes}.
% , defined in 
% a similar way via the intersections with the maximum cliques in $\HH{n}$.
\end{remark}

\begin{remark}\label{r:latin}
 \emph{Latin unitrades} 
 (\emph{latin bitrades}, \emph{latin hypercubes}) 
 are defined similarly to the 
 extended $1$-perfect unitrades, 
 via the intersections with the maximum cliques, but in the Hamming graph $H(n,q)$,
 instead of the halved $n$-cube $\HH{n}$. 
 The latin hypercubes (in the case $n=3$, the latin squares) are known in coding theory as the unrestricted (not necessarily linear or additive) 
 distance-$2$ MDS codes. 
 There are constructions of binary $1$-perfect codes 
 from latin hypercubes \cite{Phelps84}, \cite{Romanov:concat},
 which can be adopted to constructions of $1$-perfect bitrades 
 from latin bitrades \cite{Kro:small:2017} and, similarly,
 of $1$-perfect unitrades from latin unitrades. 
\end{remark}

\begin{remark}\label{r:steiner}
% The \emph{Johnson graph} $J(n,k)$ is a subgraph of $\HH{n}$ induced by 
% the weight-$k$ vertices (in the case of $k$ odd, $\HH{n}$ is treated as a graph
% on the odd-weight binary $n$-words).
 \emph{Steiner $(n,k,k-1)$ unitrades} (\emph{Steiner $(n,k,k-1)$ (bi)trades}, \emph{Steiner systems} $S(n,k,k-1)$) 
 can be defined similarly to the 
 extended $1$-perfect unitrades, 
 but the unitrade itself consists of words of weight $k$ only,
 while the balls $B$ in the definition are centered in the words of weight $k-1$.
%  via the intersections with the maximum cliques, 
%  but in the Johnson graph $J(n,k)$,
%  instead of the halved $n$-cube $\HH{n}$.
 It is not difficult to establish that a set of binary $n$-words is a Steiner
  $(n,n/2,n/2-1)$ unitrade if and only if it is a constant-weight extended $1$-perfect unitrade of length $n$ (a similar relation takes place for bitrades).
\end{remark}

The bipartiteness is a very strong property of a unitrade 
(actually bipartite unitrades of all kinds considered above 
are studied as bitrades, in the terminology popular in the theory of latin squares and latin hypercubes, 
or trades, in an alternative terminology popular in the design theory).
A bipartite $1$-perfect unitrade (as well as latin trades and Steiner trades) is a set of nonzeros 
of some $\{0,\pm 1\}$-valued eigenfunction 
of the Hamming graph (in the case of Steiner trades, the Johnson graph) see e.g. \cite{Kro:small:2017} for details.
Using algebraico-combinatorial tools, several distance invariant properties are derived for eigenfunctions of the Hamming graphs,
see e.g. \cite{Kro:struct}, \cite{Kro:interw}, \cite{Vas:2013}.
Below we discuss two important corollaries of these properties,
trying to generalize them to the class of unitrades. 
We will see that, while unitrades in general cannot be represented by eigenfunctions, in oppose to bipartite unitrades, some properties, in a weaker form, can be generalized using combinatorial approaches.

%==!==!==!==!==!==!==!==!==!==!==!==!==!==!==!==!
\subsection{The antipodality and the radius of the set}\label{s:radius}

The following property for $1$-perfect codes is well-known \cite{ShSl},
while for bipartite $1$-perfect  unitrades, it is a straightforward generalization, see e.g. \cite[Cor.~1]{Kro:small:2017}.

\begin{proposition}\label{p:antipodal}
 Every bipartite (extended) $1$-perfect unitrade $U$ in $H(n,2)$ is antipodal:
 if $x\in U$, then $x+\bar 1 \in U$, where $\bar 1$ is the all-one word and $+$ is the coordinatewise addition modulo $2$.
\end{proposition}

In other words, the \emph{inner radius} 
$\min_{x\in U}\max_{y\in U}d(x,y)$ 
of a nonempty bipartite (extended) $1$-perfect unitrade $U$ is $n$. For non-bipartite case, we can only prove that the radius is larger than $n/2$.

\begin{theorem}\label{th:OA}
 Let $U$ be a $1$-perfect unitrade or an extended $1$-perfect unitrade in $H(n,2)$, 
 let $i$ be some coordinate, $i\in\{0,\ldots,n-1\}$,
 and let $v$ be a binary word of length $n$.
 
 (i) In $U$, 
 the number of words having $0$ in the $i$th coordinate equals the number of words having $1$ in it
(in other words, the list of words from $U$ is 
an \emph{orthogonal array of strength $1$}.)
 
 (ii) An average Hamming distance from $v$ to the vertices of $U$ is $n/2$.
\end{theorem}

\begin{IEEEproof}
 Let us match any two words of $U$ differing in exactly two coordinates including $i$ or differing
 in only the $i$th coordinate. 
 As follows from the definition of (extended) unitrade, 
 every vertex $u$ from $U$ is matched to exactly one other vertex $v$ from $U$
 (indeed, the radius-$1$ ball whose center differs from $u$ in only the $i$th coordinate
  contains $u$ and exactly one other element of $U$). 
  Since every such $u$ and $v$ differ in the $i$th position, 
  (i) is proven.
  (ii) is a simple corollary of (i).
\end{IEEEproof}

So, the inner radius $\min_{x\in U}\max_{y\in U}d(x,y)$ of a (extended) $1$-perfect unitrades
is between $n/2$ and $n$. The minimum of this value remains unknown.

%==!==!==!==!==!==!==!==!==!==!==!==!==!==!==!==!
\subsection{The minimum cardinality of a unitrade}\label{s:min}

The following fact was firstly proved in \cite{Sol90en}.
Formally, the claim was stated (in different terminology) 
only for the unitrades that are symmetric difference 
of two binary $1$-perfect codes, 
but the proof is applicable to any bipartite $1$-perfect unitrade.
\begin{proposition}\label{p:min}
 The minimum cardinality of a bipartite $1$-perfect unitrade in $H(n,2)$,
 $n$ odd, is $2^{(n+1)/2}$.
 % The minimum cardinality of a bipartite extended $1$-perfect unitrade in $H(n,2)$ is $2^{n/2}$.
\end{proposition}
In the modern literature, starting from \cite{EV:94}, 
different variants of Propositions~\ref{p:min}
are usually explained utilizing the distance invariant properties 
of the bipartite $1$-perfect unitrades,
which cannot be generalized to the unrestricted case.

To prove the same fact for the non-bipartite unitrade,
we adopt the combinatorial approach from \cite{Sol90en}. 
It can be applied without any changes in the unrestricted case 
to evaluate the number of words of weight less than $n/2$ in a $1$-perfect unitrade.
However, to evaluate the number of unitrade words of weight larger than $n/2$,
the antipodality is utilized in \cite{Sol90en}. 
The non-bipartite unitrades have no antipodality in general, so we need another argument,
which occurs to be the most complicate part of the proof.

\begin{proposition}\label{p:unitrade:lb}
 The minimum cardinality of 
 a nonempty extended $1$-perfect unitrade in $H(n,2)$, $n$ even, is $2^{n/2}$.
 The minimum cardinality of 
 a nonempty $1$-perfect unitrade in $H(n,2)$, $n$ odd, is $2^{(n+1)/2}$. 
\end{proposition}
\begin{IEEEproof}
Let $U$ be an extended $1$-perfect unitrade in $H(n,2)$. 
Assume without loss of generality that $U$ contains the all-zero word.
Denote by $W$ the number of words in $U$ and by $W_i$ the number of words of
weight $i$ in $U$. Denote by $W_i^+$ ($W_i^*$, $W_i^-$)  the number of pairs $(u,v)$ of words from $U$
at distance $2$ from each other
such that $\wt(u)=i$ and $\wt(v)=i+2$ ($\wt(v)=i$, $\wt(v)=i-2$, respectively).
In particular, we have $W_0^- = W_0^* = W_n^* = W_n^+ =0$. 
From the definition of a unitrade,
it can be seen that 
\begin{equation}\label{eq:both}
2W_i^- + 2W_i^* + 2W_i^+ = nW_i,
\end{equation}
but we will prove a more detailed form of \eqref{eq:both},
splitting it into two identities \eqref{eq:down}
and \eqref{eq:up} as follows. 
Every vertex $u$ of weight $i$ belongs to $i$ radius-$1$ balls with the center 
of weight $i-1$. Each such ball has another vertex $v$ in $U$, 
with $\wt(v)=i-2$ or $\wt(v)=i$. 
The vertices $u$ and $v$ belong 
to exactly $2$ common radius-$1$ balls, 
but in the first case those two balls 
both have centers of weight $i-1$, 
while in the second case only one 
of the centers is of weight $i-1$. 
We deduce that
\begin{equation}\label{eq:down}
2 W_i^- + W_i^* = i W_i.
\end{equation}
By similar arguments,
\begin{equation}\label{eq:up}
W_i^* + 2 W_i^+ = (n-i)W_i.
\end{equation}
From the last two equations, we have 
$$
(n-i)(2 W_i^- + W_i^*) = i (W_i^* + 2 W_i^+) 
$$
\begin{IEEEeqnarray}{c}\label{eq:Wi}
 2 i W_i^+ = 2 (n-i) W_i^- + (n-2i)W_i^*, 
 \qquad i=2,4,...,n-2 .
\end{IEEEeqnarray}
Also, trivially, 
\begin{IEEEeqnarray}{c}\label{eq:Wii-2}
 W_i^- = W_{i-2}^+ , \qquad i=2,4,...,n.
\end{IEEEeqnarray}
We formally consider 
\eqref{eq:Wi}--\eqref{eq:Wii-2} 
as a system of $n-1$ linear equations
with respect to the $n-1$ variables 
$W_2^-$, $W_4^-$, \ldots, $W_n^-$,  
$W_2^+$, $W_4^+$, \ldots, $W_{n-2}^+$,
where $W_{0}^+$ and $W_i^*$, $i=2,4,...,n-2$, 
are right-side parameters.
We find the solution of the system for each basis set of parameter values (temporarily, we forget about the meaning
of the parameters and assume that they can possess any values, 
in particular,  $W_{0}^+$ can be $0$), 
also calculating the value of $W$,
$W=W(W_0^+, W_2^* , \ldots , W_{n-2}^*)=\frac 2n (W_0^+ + W_2^+ + \ldots + W_{n-2}^+  + W_2^* + \ldots + W_{n-2}^* + W_2^- + \ldots + W_n^-)$,
for each of those solutions. The solutions are easy to check by substituting the values to \eqref{eq:Wi} and \eqref{eq:Wii-2}.

(i) If $W_{0}^+=n/2$ and $W_i^*=0$, $i=2,4,...,n$, then
\begin{IEEEeqnarray*}{rCl}\label{eq:asdfjh}
 2W_i^- &=& i\binom{n/2}{i/2},
 \qquad i=2,4,...,n,\\
 2W_i^+ &=& (n-i)\binom{n/2}{i/2},
 \qquad i=0,2,...,n-2, \\
 W &=& \sum_{j=0}^{n/2}\binom{n/2}{j} =  2^{n/2}.
\end{IEEEeqnarray*}

(ii) If $W_{0}^+=0$, $W_k^*=1$ for some $k \in \{2,4,...,n-2\}$, 
and  $W_i^*=0$ for every $i \in \{2,4,...,n-2\}\backslash\{k\}$, 
then
\begin{IEEEeqnarray*}{rCl}\label{eq:vgstrd}
 2W_i^- &=& 0,
 \qquad i=2,4,...,k,\\
 2W_i^+ &=& 0,
 \qquad i=0,2,...,k-2,\\
 2W_i^- &=& \beta i\binom{n/2}{i/2},
 \qquad i=k+2,k+4,...,n, \\
 &&\beta=\frac{n-2k}{k(n-k)} \Big/ \binom{n/2}{k/2},
 \\
 2W_i^+ &=& \beta (n-i)\binom{n/2}{i/2},
 \qquad i=k,k+2,...,n, \\
 W &=& \frac 2n + \frac\beta{n} (n-k)\binom{n/2}{k/2} + \beta \sum_{j=k/2+1}^{n/2}\binom{n/2}{j}.
\end{IEEEeqnarray*}

We claim that $W>0$ for every $k$. If $k\le n/2$, then trivially we get $\beta\ge 0 $ and $W>0$.
In the case $k > n/2$, we have $\beta<0$, and the inequality 
$W>0$ is equivalent to $nW/\beta <0$, that is,
\begin{equation}\label{eq:toprove}
(n-k)\binom{n/2}{k/2} 
+ n\sum_{j=k/2+1}^{n/2}\binom{n/2}{j}
<
\frac{2k(n-k)}{2k-n}\binom{n/2}{k/2}.
\end{equation}
After substituting $n=2m$ and $k=2m-2l$, \eqref{eq:toprove}
turns to
$$
l \binom{m}{l} 
+ m\sum_{j=0}^{l-1}\binom{m}{j} 
<
\frac{2l(m-l)}{m-2l}\binom{m}{l},\qquad l<\frac{m}{2}
.
$$
After adding $(m-l)\binom{m}{l}$ to the both sides and then dividing by $m$, we get
\begin{equation}\label{eq:pot}
\sum_{j=0}^{l}\binom{m}{j}
<
\frac{m-l}{m-2l}\binom{m}{l} 
.
\end{equation}
The last inequality was proved in \cite[p.50]{Pot:2004}; for completeness, 
we repeat the proof here.

\begin{itemize}
 \item We have $m>2l$.
Hence, for any $i\ge 0$ we have $m-l+i>l$, and
\begin{equation}
{\binom{m}{l-i}}\Big/{\binom{m}{l}}
=
\prod_{s=0}^{i-1}{\frac{l-s}{m-l+i-s}}
\leq
\left(\frac{l}{m-l+i}\right)^{i}
\leq
\left(\frac{l}{m-l}\right)^{i}.
\label{eq:e4'}
\end{equation}
 Since $0<\frac{l}{m-l}<1$, we can find the sum of
 the infinite geometric series
% $\sum_{i=0}^l\gamma^i<\frac{1}{1-\gamma}$, where $\gamma=\frac{l}{m-l}$.
 \begin{equation}\label{eq:geom}
 \sum_{i=0}^\infty \left(\frac{l}{m-l}\right)^i=\frac{1}{1-\frac{l}{m-l}} 
 = \frac{m-l}{m-2l}.
\end{equation}
 Utilizing (\ref{eq:e4'}) and \eqref{eq:geom}, we get
$$
\sum_{j=0}^l {\binom{m}{j}} = \sum_{i=0}^l {\binom{m}{l-i}} 
\;\stackrel{\eqref{eq:e4'}}\leq\;
{\binom{m}{l}}\sum_{i=0}^l \left(\frac{l}{m-l}\right)^i 
\;\stackrel{\eqref{eq:geom}}{<}\;
\frac{m-l}{m-2l}{\binom{m}{l}}, $$
which validates \eqref{eq:pot} and hence~\eqref{eq:toprove}.
\end{itemize}

So, we see that $W$ grows with
the growth of any parameter $W_i^*$, $i\in\{2,4,\ldots,n-2\}$.

Since $W_0=1$, we have $W_0^+=n/2$.
Hence, the cardinality $W$ of $U$ 
equals $2^{n/2}$
if $W_i^* = 0$ for all $i$, 
and it is larger than 
$2^{n/2}$ if $W_i^*$ is positive 
for some $i$ from $\{2,4,...,n-2\}$. 
This proves the lower bound
on the cardinality 
of an extended $1$-perfect unitrade.
An example attending this bound 
is $\{(\bar x,\bar x):\ \bar x\in\{0,1\}^{n/2} \}$. 
The second claim of the proposition 
is straightforward from the first one.
\end{IEEEproof}

% [?????????? Do we need this?????????????] So, the class of unitrades are 
% a further generalization of the class of $1$-perfect codes
% that, in comparing to the class of bitrades, 
% loses many algebraic properties, 
% but still inherits some global metric characteristics of the objects, 
% which are derived from the local definition.
% While one can say that the deep study of the unitrades needs more motivation, 
% keeping in mind the existence of such generalization is very important;
% it shows that the search of a key tool to solve some problem 
% for $1$-perfect code
% (e.g., one known problem in the asymptotic of the double logarithm of the number of $1$-perfect binary codes)
% should not be restricted by tools using strong algebraic properties connected to the eigenfunctions. 

%==!==!==!==!==!==!==!==!==!==!==!==!==!==!==!==!
\subsection{Two constructions}\label{s:constr}

In this subsection, we present two constructions 
of extended $1$-perfect unitrades.
The first one (Proposition~\ref{p:conc}) is recursive 
and gives unitrades called reducible. 
The second construction (Proposition~\ref{p:uni}) 
gives one example 
of an irreducible non-bipartite 
extended $1$-perfect unitrade in every 
$H(n,2)$, $n=6,8,10,\ldots$.
It happens that with 
these two constructions 
(Propositions~\ref{p:conc} 
and~\ref{p:uni}),
one can construct all
extended $1$-perfect unitrades
in $H(6,2)$ and all
non-bipartite
extended $1$-perfect unitrades
in $H(8,2)$.

\begin{proposition}[the concatenation of unitrades]
\label{p:conc}
 If $U$ and $V$ are extended $1$-perfect unitrades in $H(m,2)$ and $H(n,2)$, respectively,
 then 
 \begin{IEEEeqnarray}{c}\label{eq:conc}
 W=\{(\bar u | \bar v):\,\bar u \in U, \bar v \in V\} 
 \end{IEEEeqnarray}
 is an extended $1$-perfect unitrade in $H(m+n,2)$.
 Moreover,
 $W$ is a bipartite extended $1$-perfect unitrade
 if and only if both $U$ and $V$
 are bipartite extended $1$-perfect unitrades.
\end{proposition}

 \begin{IEEEproof}
  Straightforward from the definitions.
 \end{IEEEproof}

We say that an extended 
$1$-perfect unitrade is \emph{reducible}
(\emph{irreducible})
if it can (cannot) be represented as a
concatenation~\eqref{eq:conc},
up to permutation of the coordinates.
% Below, the notation
% $\langle \ldots \rangle$ stands 
% for the linear span of binary words
% understood as vectors over the binary field
% $\mathrm{GF}(2)$.

\newcommand\bmodn{}
\begin{proposition}[example 
of an irreducible non-bipartite 
extended $1$-perfect unitrade]
\label{p:uni}
Let $n$ be an even number larger than $4$. 
Define 
$$
L = \langle 
\underbrace{00\ldots00\,00\,11\,11}_{\text{length }n}
,\ 
\underbrace{00\ldots00\,11\,00\,11}_{\text{length }n}
,\ 
\ \ldots \ ,\ 
\underbrace{11\,00\ldots00\,11}_{\text{length }n} 
\rangle,
$$
where
$\langle \ldots \rangle$ denoted
the linear span of the binary words
understood as vectors over the binary field
$\mathrm{GF}(2)$.
For each $i=0,2,\ldots, {n-2}$, 
let $L_i$ consist
of all words of $L$ 
with the values of 
$i$th, 
$(i+1)$th,
$(i+2\bmodn)$th, 
and $(i+3\bmodn)$th 
coordinates 
changed to $0$, $1$, $1$, $0$, 
respectively (the coordinates are calculated modulo $n$).
Then the set 
$$L_* = L_*(n)= 
L \cup L_0\cup L_2\cup 
\ldots\cup L_{n-2}$$
is an irreducible non-bipartite 
extended $1$-perfect unitrade. 
Moreover, $L_* + 0101...01$ 
is constant-weight; 
i.e., it is also a Steiner $(n,n/2,n/2-1)$ 
unitrade (see Remark~\ref{r:steiner}).
\end{proposition}

% The proof is straightforward. 
% The non-bipartiteness follows 
% by Proposition~\ref{p:antipodal} 
% from the fact that 
% the complement of any word from 
% $L_i$ is not in $L_*$.
 \begin{IEEEproof}
  For a binary word 
  $\bar x = (x_0,\ldots, x_{n-1})$, 
  we consider the pairs
  $(x_i,x_{i+1})$ 
  of the values of two subsequent coordinates, 
  starting from a coordinate 
  with an even number 
  $i=0,2,\ldots,n-2$.
  Such a pair is called \emph{zero}
  if $x_i=x_{i+1}=0$, 
  and \emph{non-zero} otherwise; 
  \emph{odd} if  $x_i+x_{i+1}=1$, 
  and \emph{even} otherwise.
  
  Let us check that $L_*$ satisfies 
  the definition of unitrade.
  The set $L$ consists 
  of all the words 
  with only even pairs 
  and even number of non-zero pairs.
  Every word $\bar y$ 
  at Hamming distance $1$ from $L$ 
  has exactly one odd pair $(y_i,y_{i+1})$. 
  Clearly, $\bar y$ is adjacent 
  to only one word of $L$, 
  with $(0,0)$ or $(1,1)$ 
  in the corresponding pair of coordinates,
  depending on the parity 
  of the number of non-zero 
  pairs in $\bar y$.
  If $(y_i,y_{i+1})=(0,1)$, 
  then $\bar y$ is adjacent 
  to one word from $L_i$, 
  with $1$ and $0$ 
  in the $(i+2\bmodn)$th 
  and $(i+3\bmodn)$th coordinates, respectively.
  If $(y_i,y_{i+1})=(1,0)$, 
  then $\bar y$ is adjacent 
  to one word from $L_{i-2\bmodn}$, 
  with $0$ and $1$ in the $(i-2\bmodn)$th 
  and $(i-1\bmodn)$th coordinates.
  It is easy to see 
  that the two words above, 
  one from $L$ and one from 
  $L_i$ or  $L_{i-2\bmodn}$, 
  are the only words from $L_*$ 
  adjacent to $\bar y$.
  
  The other group of words 
  at Hamming distance $1$ 
  from $L_*$ consists of words 
  $\bar y$
  with exactly three odd pairs. 
  Moreover, two of these pairs 
  are in four consequent 
  coordinates with numbers
  $i$, ${i+1}$, ${i+2\bmodn}$, 
  ${i+3\bmodn}$ and have 
  the values $(0,1)$, $(1,0)$, 
  respectively.
  These two pairs 
  are determined uniquely.
  Changing the values 
  in the third odd pair 
  to $(0,0)$ or $(1,1)$,
  we obtain the two 
  words of $L_*$ 
  (more tightly, of $L_i$) 
  at Hamming distance $1$ 
  from $\bar y$.
  
  So, a sphere of radius $1$ centered in a 
  word of odd weight contains $0$ or $2$ elements of $L_*$; 
  i.e., $L_*$ is an 
  extended $1$-perfect unitrade 
  by the definition.
 
  Since the complement of any word from 
  $L_i$ is not in $L_*$, 
  we see from Proposition~\ref{p:antipodal} 
  that $L_*$ cannot be bipartite.

  To ensure that $L_*$ is irreducible,
  we consider a graph on the $n$ 
  coordinates as the vertices,
  two coordinate being adjacent 
  if $L_*$ contains two words
  differing only in these coordinates.
  For every $i$ from $\{0,2,\ldots, {n-2}\}$,
  there are two words in $L$ and $L_i$, respectively,
  that differ in the $i$th and $(i+2 \bmodn)$th 
  coordinates (with the values $1,1,0,0$ and $0,1,1,0$
  in the consequent four coordinates starting from the $i$th coordinate);
  similarly,
  there are two words
  in $L$ and $L_i$,
  respectively,
  that differ in the
  $i$th and $(i+3 \bmodn)$th coordinates
  (with the values $1,1,1,1$ and $0,1,1,0$).
  It follows that the considered graph is connected,
  which obviously not the case for a reducible 
  unitrade.
  
  The last statement 
  of the proposition is obvious 
  from the following 
  three facts:
  (i)
  for every 
  $\bar x = (x_0,\ldots,x_{n-1}) \in L+0101...01$ 
  and every
  $j$ from $\{0,2,\ldots, {n-2}\}$,
  we have $\{x_j,x_{j+1}\} = \{0,1\}$;
  (ii) we have the same 
  for every $i$ from $\{0,2,\ldots, {n-2}\}$,
  every $\bar x$ from $L_i+0101...01$,
  and every $j$ from 
  $\{0,2,\ldots, {n-2}\}\backslash\{i,i+2\bmodn\}$;
  (iii) for every $i$ from $\{0,2,\ldots, {n-2}\}$
  and
  every $\bar x$ from $L_i+0101...01$,
  we have 
  $(x_i,x_{i+1},x_{i+2\bmodn},x_{i+3\bmodn})
  =(0,0,1,1)$.
  \end{IEEEproof}

\subsection{Classification of small unitrades}\label{s:small}

The bipartite extended $1$-perfect unitrades in $H(n,2)$, $n=6,8,10$, 
where classified in \cite{Kro:small:2017} in terms of bitrades (see Remark~\ref{r:perfect} for the definition). Here, we briefly describe the results of the classification of all small unitrades
up to length $10$, mainly focusing on the non-bipartite case.
The classification algorithm is described in the Appendix.

The only non-bipartite unitrade of length $6$, up to equivalence, is $L_*(6)$.
The two nonequivalent non-bipartite unitrades of length $8$
are $L_*(8)$ and $L_*(6) 10 \cup L_*(6) 01$.
There are $38$ equivalence classes of 
primary unitrades of length $10$ \cite{uni10}, 
with unitrade cardinalities
\textbf{\underline{32}}, \underline{40}, 
\textbf{\underline{48}}, 
\underline{48}, 
\underline{50}, 
\textbf{\underline{56}}, 
\underline{56}, 
\underline{56}, 
\underline{58}, 
\underline{62}, 
62, 
\textbf{64}, 
\textbf{64}, 
\textbf{64}, 
\underline{70}, 
\underline{70}, 
70, 
\textbf{\underline{72}}, 
\underline{72}, 
\underline{72}, 
72, 72, 72, 72, 72, 72, 72, 76,
\textbf{80}, 
80, 80, 80, 86, 88, 88, 96, 96, 96; eight of them
(marked by bold) are bipartite;
eleven (underlined) have constant-weight representatives. 
The unique non-bipartite antipodal unitrade of length $10$ is notable; 
it is a vertex orbit of cardinality $80$ under the automorphism group of the $(10,40,4)$ Best code 
(see \cite{ConSlo:94}).
There are only $4$ nonempty non-primary unitrades of length $10$,
of cardinality \textbf{32}+\textbf{32}, \textbf{32}+\textbf{32}, 
40+40, and 40+40 (in particular, 
we have convinced that all $2$-fold $1$-packings of size $96$ 
in $H(9,2)$ are primary unitrades).

\appendix
\section*{Classification of 
$\lfloor n/2 \rfloor$-regular
triangle-free subgraphs of the halved $n$-cube}
As follows from Corollary~\ref{cor:wd}, any binary even-weight 
$\lambda$-fold $1$-packing attaining a bound in Theorem~\ref{th:main1}
induces an $\lfloor n(\lambda-1)/2\rfloor$-regular subgraph of the halved $n$-cube.
Moreover, any subgraph corresponds to a $2$-fold $1$-packing
if and only if it has no induced triangles. 
As a result, optimal binary even-weight 
$2$-fold $1$-packings of length up to $10$ can be classified 
with the classification of $\lfloor n/2 \rfloor$-regular
triangle-free subgraphs of $\HH{n}$, $n\le 10$.
The following pseudocode describes a breadth-first algorithm
for the search all inequivalent subsets of vertices of the halved $n$-cube
that induce a connected $\lfloor n/2 \rfloor$-regular
triangle-free subgraph.
The partial solutions found at each step are validated
using the double-counting approach~\cite[\S10.2]{KO:alg} 
based on the orbit-stabilizer theorem
(in practice, the isomorph rejection 
and double-counting validation were processed only
at the first three recursive steps for partial solutions, 
and for the final solutions).

\mbox{}\\
\verb|define RECURSION(|$s$\verb|): #| \emph{$s$ is the step number} \\
\verb|    if |$T_{s-1}^{+}=T_{s}^{+}=\{\}$\verb|: |  \\
\verb|        FOUND_SOLUTION() #| \emph{record the solution $(T_0,T_1)$} \\
\verb|    else if |$T_{s-1}^{+}=\{\}$\verb|:| \\
\verb|        if |$T$\verb| is new, up to equivalence: #| \emph{isomorph rejection} \\
\verb|            RECURSION(|$s+1$\verb|) #| \emph{go to the next step} \\
\verb|    else:  | \\
\verb|        choose |$v$\verb| from |$T_{s-1}^+$ \\
\verb|        |$T_{s-1}^+:=T_{s-1}^+\backslash \{v\}$ \\ 
\verb|        for all |$\lfloor n/2\rfloor$\verb|-subsets |$N$\verb| of the neighborhood of |$v$ \\ 
\verb|          such that |$N\cup T$\verb| is triangle-free do:| \\ 
\verb|            |$N^{+}:=N\backslash T$\verb| #| \emph{new vertices to add} \\ 
\verb|            |$T_{s}^{+}:=T_{s}^{+} \cup N^+$ \\ 
\verb|            |$T:=T \cup N^+$\\ 
\verb|            RECURSION(|$s$\verb|) | \\ 
\verb|            |$T_{s}^{+}:=T_{s}^{+} \backslash N^+$ \\ 
\verb|            |$T:=T \backslash N^+$ \\ 
\verb|        |$T_{s-1}^+:=T_{s-1}^+ \cup \{v\} $ \\ 
% \verb|#| \emph{now, the main part of the algorithm} \\
$T := \{00...0 \}$ \verb| #| \emph{start from the all-zero word}
\\
$T_0^{+} : = T$ \verb| #| \emph{$T_i^{+}$ keep the chosen vertices with the ``unsolved'' neighborhood}
\\
$T_i^{+} := \{\},\quad i=1,2,\ldots$
\\
\verb|RECURSION(1)|

After the classification of connected 
subgraphs, they can be combined in all inequivalent ways to 
form a disconnected subgraphs 
(for $n\le 10$, this step is straightforward,
and the number of connected components is at most $2$).

For even $n$, 
the algorithm above finds all extended $1$-perfect unitrades.
In \cite{Kro:small:2017}, the ``bipartite'' modification 
of this algorithm was used to classify the extended $1$-perfect bitrades (essentially, bipartite unitrades).
Without the bipartiteness, 
the number of search branches 
becomes essentially larger, but still doable
for the length up to $10$. The classification took almost ten hours of computer time.

%==!==!==!==!==!==!==!==!==!==!==!==!==!==!==!==!
%==!==!==!==!==!==!==!==!==!==!==!==!==!==!==!==!
%==!==!==!==!==!==!==!==!==!==!==!==!==!==!==!==!

\section*{Acknowledgements}
The authors are grateful to the referees for the detailed reviews, which enabled us to greatly
improve the presentation of the results.

 The research was
carried out at the Sobolev Institute  of Mathematics at the expense
of the Russian Science Foundation, Grants 14-11-00555 
(Sections~\ref{s:bounds}-A, \ref{s:9}-A, ~\ref{s:uni}-CD) 
 and 18-11-00136 
 (Sections~\ref{s:q},~\ref{s:mds}, ~\ref{s:bounds}-B,
 \ref{s:9}-B, 
 and~\ref{s:uni}-AB).

% \bibliographystyle{IEEEtranS}
% \bibliography{../../k}
% \end{document}

% Generated by IEEEtranS.bst, version: 1.14 (2015/08/26)
\providecommand\href[2]{#2} \providecommand\url[1]{\href{#1}{#1}}
  \def\DOI#1{{\small {DOI}:
  \href{http://dx.doi.org/#1}{#1}}}\def\DOIURL#1#2{{\small{DOI}:
  \href{http://dx.doi.org/#2}{#1}}}

\end{document}